\documentclass[preprint,12pt]{elsarticle}

\usepackage{amssymb}
\usepackage{amsmath}

\usepackage{subcaption}
\usepackage{graphicx}
\usepackage{dcolumn}
\usepackage{bm}
\usepackage[hidelinks]{hyperref}

\usepackage{lineno}
\makeatletter%
\DeclareRobustCommand{\cev}[1]{%
  {\mathpalette\do@cev{#1}}%
}
\newcommand{\do@cev}[2]{%
  \vbox{\offinterlineskip
    \sbox\z@{$\m@th#1 x$}%
    \ialign{##\cr
      \hidewidth\reflectbox{$\m@th#1\vec{}\mkern4mu$}\hidewidth\cr
      \noalign{\kern-\ht\z@}
      $\m@th#1#2$\cr
    }%
  }%
}
\makeatother
\journal{Nuclear Instuments and Methods A}

\begin{document}
\begin{frontmatter}



\title{Relativistic beam loading, recoil-reduction, and residual-wake acceleration with a covariant retarded-potential integrator}


\author[inst1]{Benjamin Folsom}
\ead{benjamin.folsom@axis.com}
\affiliation[inst1]{organization={European Spallation Source},
            addressline={Partikelgatan 2}, 
            city={Lund},
            postcode={22484}, 
            country={Sweden}}

\author[inst1]{Emanuele Laface}

\begin{abstract}
An algorithm is demonstrated that performs first-principles tracking of relativistic charged-particles. A covariant approach is used which relies on retarded vector potentials for trajectory integration instead of performing electromagnetic field calculations. When accounting for retardation effects, the peak vector potential and corresponding Lorentz force in the direction of travel increase asymptotically for high-$\beta$ particles. This produces a very strong field distribution at small angles from the particle's direction of travel, which can result in considerable change in momentum when approaching a conducting or charged object. We study these dynamics using protons and electrons at relativistic energies passing through apertures in conducting surfaces, where substantial energy shifts are observed for particles passing within roughly 10 microns of the aperture boundary.

We also simulate breaking a test particle's line of sight with a conductor or other charged body. After this instant, the test particle continues to accelerate due to residual fields, but no longer produces an opposing force on any charged or conducting object; thus any recoil on the enclosing structure is effectively reduced. In this test, a 1\% energy gain is observed for an 85\,MeV electron traversing its reflected wake after having conducting plate in its path screened by a dielectric object.

We then incorporate a micro-scale dielectric laser acceleration (DLA) device into our simulations. Compared with a 2\,mm DLA on its own, we find a factor of two increase in energy gain when adding a series of conducting-surface choppers.

\end{abstract}


\begin{highlights}
\item Use of a covariant, retarded-potential integrator to track relativistic electrons and protons from first principles.
\item Acceleration demonstrated through residual fields from source charges which are screened from line-of-sight.
\end{highlights}


\end{frontmatter}


\section{\label{sec:Introduction}Introduction}

The covariant Lorentz forces can be derived from the electric and magnetic field form of the retarded Liénard--Wiechert potentials for a charged particle~\cite{jackson2012classical}
\begin{equation}
\begin{gathered}
\mathbf{B}=[\mathbf{n} \times \mathbf{E}]_{\mathrm{ret}} \\
\mathbf{E}=e\left[\frac{\mathbf{n}-\boldsymbol{\beta}}{\gamma^{2}(1-\boldsymbol{\beta} \cdot \mathbf{n})^{3} R^{2}}\right]_{\mathrm{ret}}+\frac{e}{c}\left[\frac{\mathbf{n} \times\{(\mathbf{n}-\boldsymbol{\beta}) \times \dot{\boldsymbol{\beta}}\}}{(1-\boldsymbol{\beta} \cdot \mathbf{n})^{3} R}\right]_{\mathrm{ret}}
\label{LW_forces_jackson}
\end{gathered}
\end{equation}
\vspace{-0.2cm}

\noindent Here, $\mathbf{n}$ is a unit vector pointing from a source particle toward an observation point; $R$ is the distance between the source and observation point; $\boldsymbol{\beta}$ and $\gamma$ are the relativistic velocity three vector ($\textbf{v}/c$) and Lorentz factor, respectively; $\dot{\boldsymbol{\beta}}$ is the acceleration in terms of $\boldsymbol{\beta}$; and $e$ and $c$ are the particle charge and speed of light, respectively. Note also that all terms are evaluated at retarded time $t = R / c$ prior to the present trajectory step, and Gaussian units are used throughout the article. When we refer to a test particle, this means that it is positioned at the observation point, where it ``witnesses'' the source particle. In other words, the charge and all of the $\beta$-dependent factors in Eq.~(\ref{LW_forces_jackson}) belong to the source particle (i.e. \textit{not} the particle whose dynamics will be affected at the observation point).

The theory and simulation of beam loading in cavities is well-understood~\cite{wolski2014beam}, where the usual approach is to track a particle's averaged Lorentz force on the cavity walls, so that $\boldsymbol{n}$ and $\boldsymbol{\beta}$ can be taken as approximately perpendicular such that
\begin{equation}
    \boldsymbol{E_\perp} \propto e\frac{1}{\gamma^2 R^2}
\end{equation}
\noindent Although the $1/R^2$ terms are suppressed on average for high-$\beta$ particles by the large $1/\gamma^2$ factor, there is a notable exception:

For a charged particle moving at a constant high-$\beta$ velocity, or being accelerated along its predominant $\boldsymbol{\beta}$ vector component, the magnetic and acceleration-dependent terms for an observation point on-axis with $\boldsymbol{\beta}$ (i.e. in the direction of $\mathbf{n}$) become negligible, and Eq.~(\ref{LW_forces_jackson}) can be approximated as

\begin{align}
\label{eq:E_n_headon}
\mathbf{E}_n&=e\left[\frac{\mathbf{n}-\boldsymbol{\beta}}{\gamma^{2}(1-\boldsymbol{\beta} \cdot \mathbf{n})^{3} R^{2}}\right] \underset{\beta \to 1}{\approx}  e \displaystyle\frac{\gamma^2}{2 R^2} \mathbf{n} \, ,~~ \boldsymbol{\beta} \cdot \boldsymbol{n} \to \beta^2
\nonumber \\[1.5ex]
&=e\left[\frac{(1-\beta_n)\mathbf{n}}{\gamma^{2}(1-\beta_n)^{3} R^{2}}\right] \, ,~~ \boldsymbol{\beta} \cdot \boldsymbol{n} \to |\beta|
\nonumber \\
&\approx e\left[\frac{(1+\beta)}{(1-\beta) R^{2}}\right]\mathbf{n} 
\nonumber \\
&\underset{\beta \to 1}{\approx} e\left[\frac{4 \gamma^2}{R^{2}}\right]\mathbf{n}
\end{align}

\noindent where we have dropped the ``ret'' subscript for brevity. For the approximation in the last line we treat transverse components as negligible such that $\boldsymbol{\beta}\approx\beta_n$, representing a near-perfect alignment between the source particle's velocity and the unit vector directed at the observation point. The approximation in the first line shows the case of a slight misalignment where $\boldsymbol{\beta} \cdot \boldsymbol{n} = |\beta|\operatorname{cos}\theta \approx \beta^2$. Given these limits, it is clear that this $\gamma^2$ dependence requires highly precise alignment for ultrarelativistic cases, and rapidly reduces to a $1/\gamma^2$ dependence for $\operatorname{cos}\theta > | \beta|$ (i.e. in the small-angle approximation, $\theta \lesssim \sqrt{2 (1-\beta)}$ for an overall $\gamma^2$ dependence). Thus, in this case the total Lorentz force for a charged particle at the observation point can be approximated as $e_o\mathbf{E}_n$, where $e_o$ is the observer charge.

\begin{figure}
    \includegraphics[width=0.84\textwidth,trim={0 0.65cm 0.5cm 0.5cm},clip]{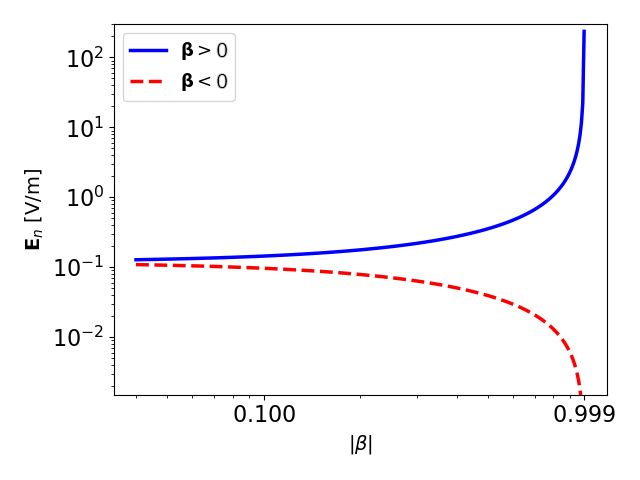}
  \caption{Electric field strengths from a charged particle approaching head-on ($\boldsymbol{\beta}>0$) or departing antiparallel ($\boldsymbol{\beta}<0$) to an observer at $R=0.01$\,mm.}
  \label{fig:rel_lortz_force_both}
\end{figure}

Figure~\ref{fig:rel_lortz_force_both} illustrates the respective asymptotic increase and decrease of such fields for high $\beta$ particles either approaching or departing from an observation point with perfect alignment. For the case of an approaching particle ($\boldsymbol{\beta}>0$) as particle velocities exceed $\gamma\,{\approx}${\,}20, the distances $R$ for which $\mathbf{E}_n$ is non-negligible become comparable to those typical of cavity or cavity-cell lengths. In typical situations, this highly peaked energy distribution is directed toward cavity exit apertures, indicating that for high-energy beams an analysis of radiated power near such elements is warranted.


We first lay the theoretical groundwork for such studies, introducing a simulation framework and then providing an qualitative discussion of secondary effects such as radiation pressure and the radiation-reaction force, along with a pedagogical model examining acceleration and recoil effects due to residual fields (i.e. where a test particle loses line of sight with a charged or conducting object).

This is followed with simulation results analyzing a benchmark case using proton collisions with heavy ions to compare our numerical model with the approximations in Eq.~(\ref{eq:E_n_headon}). We then model the energy shift and power deposition of high-$\beta$ test particles passing through narrow conducting apertures. 
It is shown that rapidly withdrawing such a surface (i.e. opening a conducting iris or chopper located at the cavity exit) allows the test particles to continue accelerating in the residual fields while reducing the longitudinal recoil on the system.

Building from these results, an additional scenario is modeled where a low-$\beta$ but highly populated ``driving'' bunch passes transversely to the beam axis behind a fixed aperture in a dielectric plate such that the oncoming high-$\beta$ test bunch travels through other particle's wake. The dielectric plate allows for line of sight to be abruptly broken between the test particle and driving bunch, similarly to the rapidly disappearing conducting surface of the previous scenario. 

Finally, we compare the acceleration of the models developed here with recent studies in dielectric-laser acceleration using an inverse Cherenkov radiation effect (ICR-DLA)~\cite{liu_2020}. For this arrangement, the GeV/m scale acceleration of ICR-DLAs is increased by a factor of two when adding cascading iris/chopper structures of roughly 2\,mm in length.

\section{Integrator}
Since all terms in Eq.~(\ref{LW_forces_jackson}) must be tracked in order to determine off-axis field contributions (i.e. to account for forces where $\boldsymbol{\beta}\cdot\mathbf{n}\not\approx 1$) we use a fully covariant tracking algorithm similar to one derived in previous works~\cite{folsom:ipac2021-tupab218,laface_2020_covariant} following the formalism of~\cite{jackson2012classical} and \cite{bordovitsyn_technique_2003} where

\begin{equation}
A^{\alpha}(x^\alpha)=\left.\frac{\vec{e}~V^{\alpha}(\tau)}{V \cdot[x-r(\tau)]}\right|_{\tau=\tau_{0}}
\label{LW_pots}
\end{equation}

\noindent are the Liénard--Wiechart potentials and $\vec{e}$ is source charge, $x^\alpha$ is the observer position, and $r^\alpha(\tau)$ is the source charge position. To avoid confusion with contravariant vectors' subscripts, we use the typical arrow accent to label the ``source'' parameters where necessary, and a reversed left-pointing arrow for ``observer'' parameters (for example: the observer charge becomes $e_o\equiv\cev{e}$).\footnote{These accents do \textit{not} designate 3-vectors, which are set in boldface type.}

The source particle's four-position and four-velocity are defined in the usual way:
\begin{align}
    r^\alpha(\tau) &= \{c\tau,\boldsymbol{r}\}
    \nonumber \\
    V^\alpha(\tau) &= \{c\gamma,\gamma\boldsymbol{u}\}
\end{align}
where $\tau$ is the proper time for the observer point, \textbf{u} is the velocity three-vector, and the metric $g^{\alpha \beta}=\{1, -1, -1, -1\}$ is used throughout. The retarded time $\tau_0$ is then defined using the light-cone constraint
\begin{align*}
    \left[x-r(\tau_0)\right]^2 = 0
\end{align*}

As with Eq.~(\ref{LW_forces_jackson}), the condition $\tau=\tau_0$ is used when evaluating any instance of the retarded potential or its resulting fields. This constraint is also used to define $R\equiv x_0-r_0(\tau_0)=|\boldsymbol{x}-\boldsymbol{r}(\tau_0)|$ along with $R^\rho=\{R,\textbf{n} R\}$. Thus, the denominator of Eq.~(\ref{LW_pots}) reduces to
\begin{equation}
    V\cdot [x-r] = V^\rho R_\rho =  \gamma c R(1-\boldsymbol{\beta}\cdot \textbf{n})
\end{equation}
recalling that $\textbf{n}$ is the unit vector pointing from the source to the observer.

The covariant equations of motion for an observer particle can be defined using the conjugate momentum
\begin{equation}
    \mathcal{P}^\alpha = \cev{m} \cev{V}^\alpha + \frac{\cev{e}}{c}A^\alpha 
    \label{mom_conj}
\end{equation}

\noindent where $\cev{m}$ and $\cev{V}^\alpha$ are the observer's mass and velocity, respectively. The Hamiltonian for a charged particle in an external field is then~\cite{jackson2012classical,barut1980electrodynamics}

\begin{align}
    H&=\frac{1}{\cev{m}}\left(P_{\alpha}-\frac{e A_{\alpha}}{c}\right)\left(P^{\alpha}-\frac{e A^{\alpha}}{c}\right)
\nonumber \\
&~~-c \sqrt{\left(P_{\alpha}-\frac{e}{c} A_{\alpha}\right)\left(P^{\alpha}-\frac{e}{c} A^{\alpha}\right)}
\end{align}

\noindent which yields the following equations of motion for conjugate momentum:

\begin{align}
\frac{d\mathcal{P}^\alpha}{d\tau}&= -\frac{\partial H}{\partial x} = \frac{\cev{e}}{\cev{m} c} \left(\mathcal{P}_{\beta}-\frac{\cev{e}}{c} A_{\beta}\right) \partial^{\alpha} A^{\beta}
\nonumber \\
&= \frac{\cev{e}}{c} \cev{V}_{\beta} \partial^{\alpha} A^{\beta}
\nonumber \\
&=\cev{e}\cev{V}_\beta\,\vec{e} \left[
\frac{\dot{\vec{V}}_{\tau}^{\beta
} R^{\alpha} - \vec{V}^\alpha \vec{V}^\beta }{\kappa^{2} \vec{\gamma}^2 R^2 c^{3}}
+\frac{R^\alpha \vec{V}^\beta }{\kappa^3 \vec{\gamma}^3 R^3 c^2} - \frac{ \dot{\vec{V}}^{\alpha}_\tau \vec{V}^{\beta} R^{\alpha} R_{\alpha}}{\kappa^3  \vec{\gamma}^3 R^3 c^4}
\right] \nonumber \\
\label{EOM_mom_nospin}
\end{align}

\noindent where we have used the shorthand 
\begin{equation*}
\kappa \equiv (1-\boldsymbol{\beta}\cdot \vec{\boldsymbol{n}})
\end{equation*}

A number of identities are helpful both in finding an expression for $\partial^{\alpha} A^{\beta}$ and determining component-wise expressions for Eq.~(\ref{EOM_mom_nospin})~\cite{jackson2012classical,bordovitsyn_technique_2003}
\begin{align}\label{velocity_identities}
\frac{d V^{\alpha}}{d \tau}&=\left[c \gamma^{4} \boldsymbol{\beta} \cdot \dot{\boldsymbol{\beta}}, c \gamma^{2} \dot{\boldsymbol{\beta}}+c \gamma^{4} \boldsymbol{\beta}(\boldsymbol{\beta} \cdot \dot{\boldsymbol{\beta}})\right]\equiv \dot{V}_\tau
\nonumber\\
\frac{d}{d\tau} \left[  V^{\rho} R_{\rho} \right] &= -c^2 + R_\rho \dot{V}_\tau^\rho
\nonumber\\
{V}^\rho V_\rho &= c^2
\end{align}

\noindent where $\dot{\beta}$ is the derivative with respect to lab time, as opposed to $\dot{V}_\tau$.
\noindent For constructing an integrator, it should be emphasized that all $\beta$-indexed vectors must be contracted while all $\alpha$-indexed ones must be taken compenent-wise. For example, the identity $R^{\rho} R_{\rho} = 0 $ should \textit{not} be used with Eq.~(\ref{EOM_mom_nospin}), since it has nonzero contributions to the individual components of $\mathcal{P}^{\alpha}$.

As an example, the final expression for the zero-index component of Eq.~(\ref{EOM_mom_nospin}) reduces to 
\begin{align}
    \frac{d \mathcal{P}^0}{d \tau} &= 
    \cev{e}  \vec{e} \, \cev{V}_{\beta} \left[ 
    \frac{\dot{\vec{V}}^\beta_\tau} {\kappa^{2} c^{3} \vec{\gamma}^{2} R}
    \nonumber - \frac{ \vec{V}^{\beta}} {\kappa^{2} c^{2} \vec{\gamma} R^{2}} \right.
    \noindent \\
    & \left. -\frac{\boldsymbol{\beta}\cdot\dot{\boldsymbol{\beta}} \vec{\gamma} \vec{V}^{\beta}}{\kappa^3 c^{3} R} 
    +  \frac{\vec{V}^{\beta} }{\kappa^3 c^{2} \vec{\gamma}^3 R^{2}}
    \right]
    \label{eq:eoms_mom_trans}
\end{align}

\noindent where the $\boldsymbol{\beta}$ and $\dot{\boldsymbol{\beta}}$ factors belong to the source particle (whose trajectory is not being calculated). These factors are separable using the first identity in Eq.~(\ref{velocity_identities}) because they match the $\alpha$ index of the conjugate momentum (whereas the $\dot{\vec{V}}^{\beta}_\tau$ factor contracts with $\cev{V}_{\beta}$). The spatial components reduce in a similar manner:
\begin{align}
    \frac{d \mathcal{P}^z}{d \tau} &= 
    \cev{e}  \vec{e} \, \cev{V}_{\beta} \left[ 
    \frac{\dot{\vec{V}}^\beta_\tau \hat{n}^{z}} {\kappa^2 c^{3} \vec{\gamma}^{2} R}
    \nonumber - \frac{ \vec{V}^{\beta} \vec{\beta}^{z}} {\kappa^2 c^{2} \vec{\gamma} R^{2}} \right.
    \noindent \\
    &- \left. \frac{ \vec{V}^{\beta} (\hat{n}^z)^2 \left(\dot{\vec{\beta}}^z + \vec{\beta}^z \vec{\boldsymbol{\beta}}\cdot\dot{\vec{\boldsymbol{\beta}}} \vec{\gamma}^{2}\right)}{\kappa^3 c^{3} R \vec{\gamma}} 
    +  \frac{\vec{V}^{\beta} \hat{n}^{z} }{\kappa^3 c^{2} \vec{\gamma}^{3} R^{2}}
    \right]
    \label{eq:eoms_mom_long}
\end{align}

The equations of motion for the position coordinates are more straightforward:
\begin{equation}
\frac{d x^{\alpha}}{d \tau}= \frac{\partial H}{\partial \mathcal{P}} = \frac{1}{\cev{m}}\left(\mathcal{P}^{\alpha}-\frac{\cev{e}}{c} A^{\alpha}\right) 
\label{eq:eoms_pos}
\end{equation}

When updating a particle's velocity and acceleration components at the end of an integration step, care should be taken to first convert the conjugate momentum values back into real momentum values using Eq.~(\ref{mom_conj}). One should not that proper time $\tau$ remains constant as the defined integrator step size, while the $\boldsymbol{\dot{\beta}}$ dependent factors are dependent on lab-frame time (see \cite{laface_2020_covariant} for a discussion on the handling of proper time as an independent variable in a similar context).

This formalism bypasses analyzing the Lorentz forces in terms of field distributions, and is practical for cases where each field source is a point particle with a known trajectory. In other words, in typical contexts field-dependent approaches (i.e. Poisson solvers) are robust and well-understood and the calculation of retarded fields may be more intuitive conceptually (for examples, see~\cite{Ryne:IPAC2018-THPAK044,ryne:ipac12-tuppp036,quattromini_LW_2009}). However, in terms of computing trajectories between interacting particles, working directly with the $A^{\alpha}$ potentials is more straightforward.

Thus, such an integrator is well-suited modeling a high-$\beta$ test charge approaching another charged ``driving'' bunch as mentioned above as a screened source. For the other models involving conducting surfaces, we use an image charge to track reflected fields from the oncoming test charge. It is highly non-trivial to assert that the Liénard--Wiechert potentials can be used to model such scenarios, especially considering surface effects, so we must assume these surfaces to be a perfectly reflecting conductors (e.g. a superconducting film with negligible defects).

\section{Model Examination}\label{model_verification}
Before analyzing results using the integrator proposed above, we consider a simple qualitative case to develop an intuition of the forces acting on any test particles and the enclosing system, particularly the recoil reduction effect. We then check that any forces which may arise in such a simulation are accounted for properly on the scales relevant to modern particle accelerators. 

\subsection{Screened source recoil reduction: A simple test case }\label{sec:recoil_reduct_pedagog}
Although the examination of energy shift and power deposition with a retarded covariant integrator is relatively straightforward, the recoil reduction due to a rapidly disappearing or screened source of charge is a more complicated effect. We provide a basic conceptual model here to clarify the concept.

The cavity for this model is taken as a perfectly absorbing blackbody, with two openings aligned along a beam axis. We denote the right-hand direction along the beam axis as $+\hat{z}$.
Two oncoming, oppositely charged, high-$\beta$ particles approach the cavity from opposite directions along $\hat{z}$ with equal-magnitude velocities and equal distances to the cavity center. We label these $\rho_L$ and $\rho_{R}$ for the particles entering from the left and right side of the cavity, respectively.

Instead of entering the cavity, the openings are closed just before the particles enter. Thus, both particles collide with the now-closed apertures, depositing their kinetic energy at the same instant in opposite directions for a zero change in net momentum.

\begin{figure}[htbp]
  \begin{subfigure}{\columnwidth}
    \centering
    \includegraphics[width=0.8\linewidth]{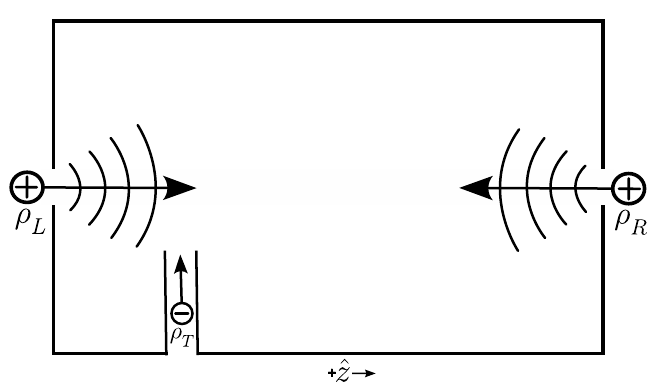}
    \caption{}
    \label{fig:model_setup}
  \end{subfigure}
  
  \begin{subfigure}{\columnwidth}
    \centering
    \includegraphics[width=0.8\linewidth]{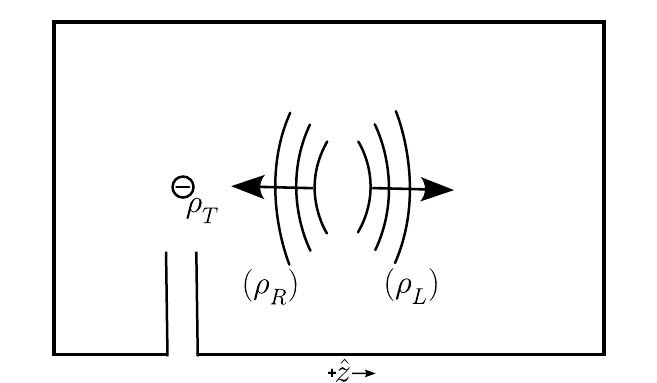}
    \caption{}
    \label{fig:model_residuals}
  \end{subfigure}
  \vspace{-1\baselineskip}
  \caption{(a) Two positively charged particles $\rho_L$ and  $\rho_R$ approach a blackbody cavity with relativistic velocities in opposite directions. A negatively charged test particle $\rho_T$ is injected through a dielectric channel with negligible energy. (b)~The two entrances are closed shortly before the particles enter the cavity, and the test particle's arrival is timed such that it only interacts with the residual field from $\rho_R$, denoted as ($\rho_R$).}
  \label{fig:model_cavity}
\end{figure}

However, the residual fields from the two particles continue to propagate through the cavity. No further changes occur to the system apart from these fields depositing their energy into the blackbody cavity walls as heat and radiation pressure. Again, these are aligned in opposite directions with respect to the beam axis and no recoil effects are observed. 

Moving beyond this benchmark case, we consider injecting a test particle, which we will denote as $\rho_{T}$. The injection apparatus is assumed to be perfectly dielectric; it is placed a short distance from the interior wall of the cavity's left-hand side, and aligned so that it can inject a test particle onto the beam axis with negligible momentum. We then take time $t_0$ to be the instant where the apertures are shut and time $t_1=t_0+\epsilon$ to be the instant where the test particle is emitted, such that the fields emitted from $\rho_L$ have traversed beyond the injection point. Thus, $\rho_T$ only interacts with the incoming fields emitted from $\rho_R$. This scenario is illustrated in Fig.~\ref{fig:model_cavity}.

The key result is that $\rho_T$ undergoes a small, but non-negligible acceleration in the $+\hat{z}$ direction and ultimately collides with the cavity wall on the right-hand side, for a net $+\hat{z}$ momentum gain to the system (having redirected momentum from $\rho_R$'s residual field into the $+\hat{z}$ direction). This is what we refer to as a screened-source recoil reduction -- a net, one-dimensional momentum shift arising not from direct, static interactions between charged bodies, but between a single charged body and residual fields from sources that are screened or otherwise neutralized. In other words, as $\rho_T$ accelerates, its resulting fields reach no charged or conducting body and thus cause no restoring force in the $-\hat{z}$ direction.

For most cases, even highly relativistic ones, this momentum shift is insignificant. However, as discussed above and highlighted in Fig.~\ref{fig:rel_lortz_force_both}, there is an asymptotic growth for velocity-dependent fields (i.e.~non-radiative fields) for head-on, high-$\beta$ interactions. In these circumstances, such a momentum shift becomes more appreciable. The case of a high-$\beta$ particle approaching a perfectly conducting surface is particularly relevant here, since the particle remains perfectly aligned with its image charge (or nearly perfectly, when accounting for retardation effects). This topic is addressed further in Section~\ref{sec:results_conducting_surface}.

\subsection{Cross-checking relevant forces on the modeled system}
For completeness, we also consider any other forces which may come into play in the pedagogical model discussed in the preceding section. 
\subsubsection*{Radiation pressure}
To examine the case where $\rho_L$ and $\rho_R$ are accelerating toward one another before colliding with their respective apertures, we should check whether the radiation pressure exerted by their residual fields is negligible with respect to the momentum gain for $\rho_T$. To do so, we can draw from Jackson's formalism~\cite{jackson2012classical}, defining the momentum shift on the test particle due to radiation pressure as
\begin{align}
    \frac{dP_{RP}}{dt} = \frac{P_R}{c} \frac{a_T}{A_R}
\end{align}
where $P_R$ is the radiated power from particle $\rho_R$, $a_T$ is the effective cross-sectional area of the test particle, and $A_R$ is the cross-sectional area of the radiated power. This can be expanded to  
\begin{align}
    \frac{dP_{RP}}{dt} = \left[\frac{2}{3}\frac{e_s^2}{m_s c^4} \left(\frac{d P_R}{dt}\right)^2\right] \cdot \frac{\pi r^2}{\Omega R^2}
\end{align}

\noindent where $P_R$ is the momentum of particle $\rho_R$, $r$ is the test particle's radius, $\Omega$ is the solid angle of the the radiated power from $\rho_R$, and $R$ is the distance between the two particles at the retarded time, as defined above.

The force from this radiation is negligible compared with the Coulomb force from the test particle's acceleration by several orders of magnitude for typical cases. Still, noting the nonlinear growth in Eq.~(\ref{eq:E_n_headon}), we should avoid the presumption that radiation effects can be ignored for extreme cases (e.g.~electrons with $\gamma>10000$ or for cavity lengths of 10\,mm or less).

Although radiation is perpendicular to the direction of travel for non-relativistic charged particles, it becomes highly forward-biased for relativistic ones, forming a double-peaked distribution with the proportionalities

\vspace{-0.2cm}
\begin{align}
    \label{radiation_max}
    \theta_{max}&\rightarrow \frac{1}{2 \gamma} 
    \\
    \theta_{rms}&\rightarrow \frac{1}{\gamma}
\end{align}

\noindent where $\theta_{max}$ and $\theta_{rms}$ are the distances between peaks and root-mean-squared angular spread of radiated power, respectively.

This allows for easy estimation of the solid angle of radiated power from~$\rho_T$. One can then calculate the change in conjugate momentum $d\mathcal{P}_T / dt$ using Eq.~(\ref{EOM_mom_nospin}), noting that the relation between proper time $\tau$ and lab-frame time $t$ is $\frac{dt}{d \tau}=\gamma$. The resulting effect is at least tens of orders of magnitude less than the change in momentum of~$\rho_T$, and can thus be ignored (even for heavy ions with a comparatively large surface area, or $d\mathcal{P}_T / dt$ values representing the upper limit of conventional accelerator capabilities).
\subsubsection*{Radiation reaction force}
Having determined the radiation pressure exerted by the incoming EM wave as negligible, there is another phenomenon needing consideration. This is the radiation reaction force, which is effectively a damping or self-recoil of a charged particle as it is accelerated by an external field and, in turn, releases radiation.

To simulate this force, we use Medina's approach~\cite{medinaRadiationReactionClassical2006} which is based on the well-known Lorentz--Abraham--Dirac equation and can be written as
\begin{equation}
    F^{RAD} = \frac{2}{3}\frac{e^2}{mc^3}\left[ \frac{d\gamma}{dt}F_{ext}-\frac{\gamma^3}{c^2}\left( F_{ext}\cdot a) \right)v \right]
\end{equation}
In this equation, $m$ is not the typical rest mass, but an electrodynamic or ``dressed'' mass, defined as $m=m_0+ U_e/c^2$ where $U_e$ is the electrostatic energy and $m_0$ is rest mass.\,\footnote{There are theoretical difficulties in choosing an appropriate electrostatic energy and, in turn, particle radius. Such concerns can be ignored so long as any length scales considered in a given problem are considerably greater than the radius of the particle. For further reading, see \cite{rafelski_2022_rad_reaction, rohrlichDynamicsChargedParticle2008,martin-lunaNovelReactionForce2022,griffithsAbrahamLorentzLandau2010a,steaneReducedorderAbrahamLorentzDiracEquation2015} along with some experimental studies complementing the topic \cite{hammondRadiationReactionUltrahigh2010,liInducedCurrentDue2022,sampathExtremelyDenseGammaRay2021}.} This model for radiation reaction force can be easily incorporated into a tracking algorithm based on Eq.~(\ref{EOM_mom_nospin}). In general, it is negligible compared with the Lorentz forces due to a high-$\beta$ particle passing through an aperture in a conducting surface or the forces associated with screened-source recoil reduction.

However, tracking this reaction force is useful in the cases where charged particles ultimately collide with a conducting surface: in the final time steps before an impending collision, accounting for the radiation reaction force prevents runaway solutions that would otherwise occur between the incoming particle and its image. Thus, the particle's approach can be probed for a final total energy preceding collision to a precision on the order of its Compton wavelength. For improved precision, the Eliezer--Ford--O'Connell formulation of this force is also worth considering~\cite{rafelski_2022_rad_reaction}.\footnote{A more robust covariant retarded-potential integrator would also incorporate magnetic-dipole moment contributions, as discussed in a prior work~\cite{folsom:ipac2021-tupab218} (see also \cite{steinmetzMagneticDipoleMoment2019a}).}
\vspace{0.2cm}
\begin{figure}[ht!]
    \centering
    \includegraphics[width=0.84\textwidth,trim={0 0.5cm 0 0.5cm},clip]{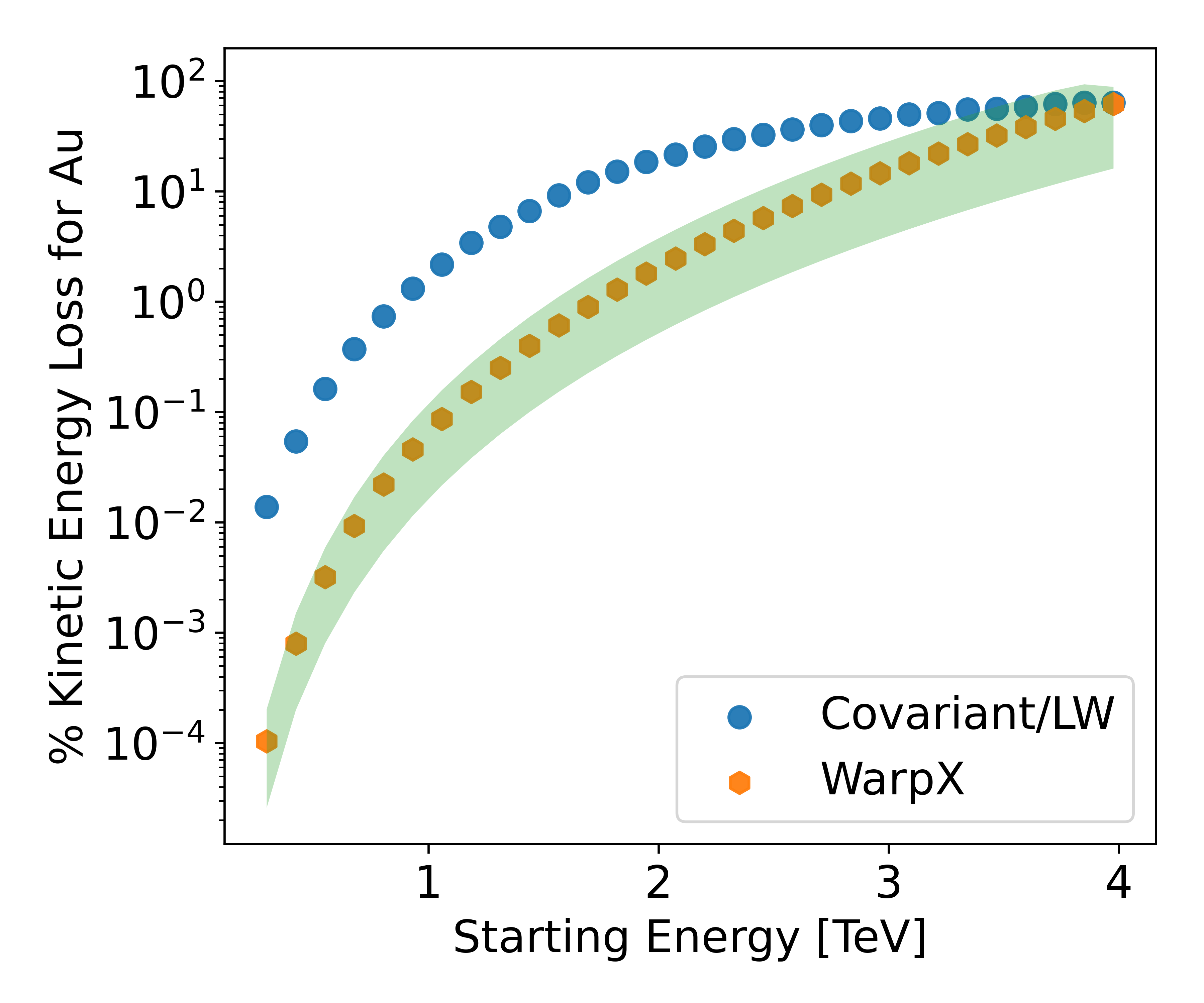}
    \caption{Comparison of this work's covariant, retarded integrator and \textsc{Warp-X} in simulating kinetic energy loss for a single Au$^{79+}$ in near-collisions with a single proton at distance of closest approach $d_T = 300\,\mathrm{nm}$. For each trial, the protons and ions have equal starting energies and traverse a 100\,mm drift from opposite ends. The \textsc{Warp-X} curve uses static fields scaled by $\gamma^2/2$, with the shaded upper and lower bounds scaled by $\gamma^2$ and $\gamma^2/4$, respectively.}
\label{fig:pb_Au}
\end{figure}

\section{RESULTS} \label{results}

\subsection{Benchmarking with Proton--Heavy Ion Interactions}

We first wish to verify that the equations of motion Eqs.~(\ref{EOM_mom_nospin}) produce numerical results comparable to the analytical approximation from Eq.~(\ref{eq:E_n_headon}), that $E_n \approx e \gamma^2 \boldsymbol{n} / (2 R^2) $ for highly aligned cases.



To do so, we implement this approximation into the tracking code \textsc{Warp-X}~\cite{VAY2018476} by including a linear factor of $\gamma^2 / 2$ when calculating the fields emitted by each particle. With this software, we selected the Lorentz invariant Vay integrator, which is is appropriate for ultra relativistic systems but does not treat the charged-particles emitted fields as velocity-dependent so that they scale nominally as $e/R^2$. 

Results for this test are shown in Fig.~\ref{fig:pb_Au}, where the stopping power of protons is shown to increase dramatically in the TeV range. In this case, each data point represents a trial where a single proton and a highly stripped gold ion (Au$^{79}$) make a near-collision at equal energies at the center of a 100\,mm drift and with a closest approach distance $d_T = 300$\,nm. For \textsc{Warp-X}, the $\gamma^2 / 2$ trials are bounded by results using multipliers of $\gamma^2$ and $\gamma^2 /4$ for the upper and lower limits, respectively. For the covariant simulation, there is little interaction between the particles after crossing near the midpoint of the drift, which is consistent with the rapid dropoff for departing particles shown in Fig.~\ref{fig:rel_lortz_force_both} for $\boldsymbol{\beta}<0$. The values for the ion's maximum kinetic energy loss \textsc{Warp-X} are thus taken from the timesteps just preceding the crossing point, in order to exclude an unrealistic isotropic field exchange as the particles depart.

While it is notable to observe such a drastic kinetic energy shift for interactions involving no nuclear forces, it should be stressed that this is a scenario modeling alignment precisions between the two particles that become increasingly stringent for higher energies.

\begin{figure}[ht!]
    \centering
    \includegraphics[width=0.84\textwidth,trim={0 0.7cm 0 0.5cm},clip]{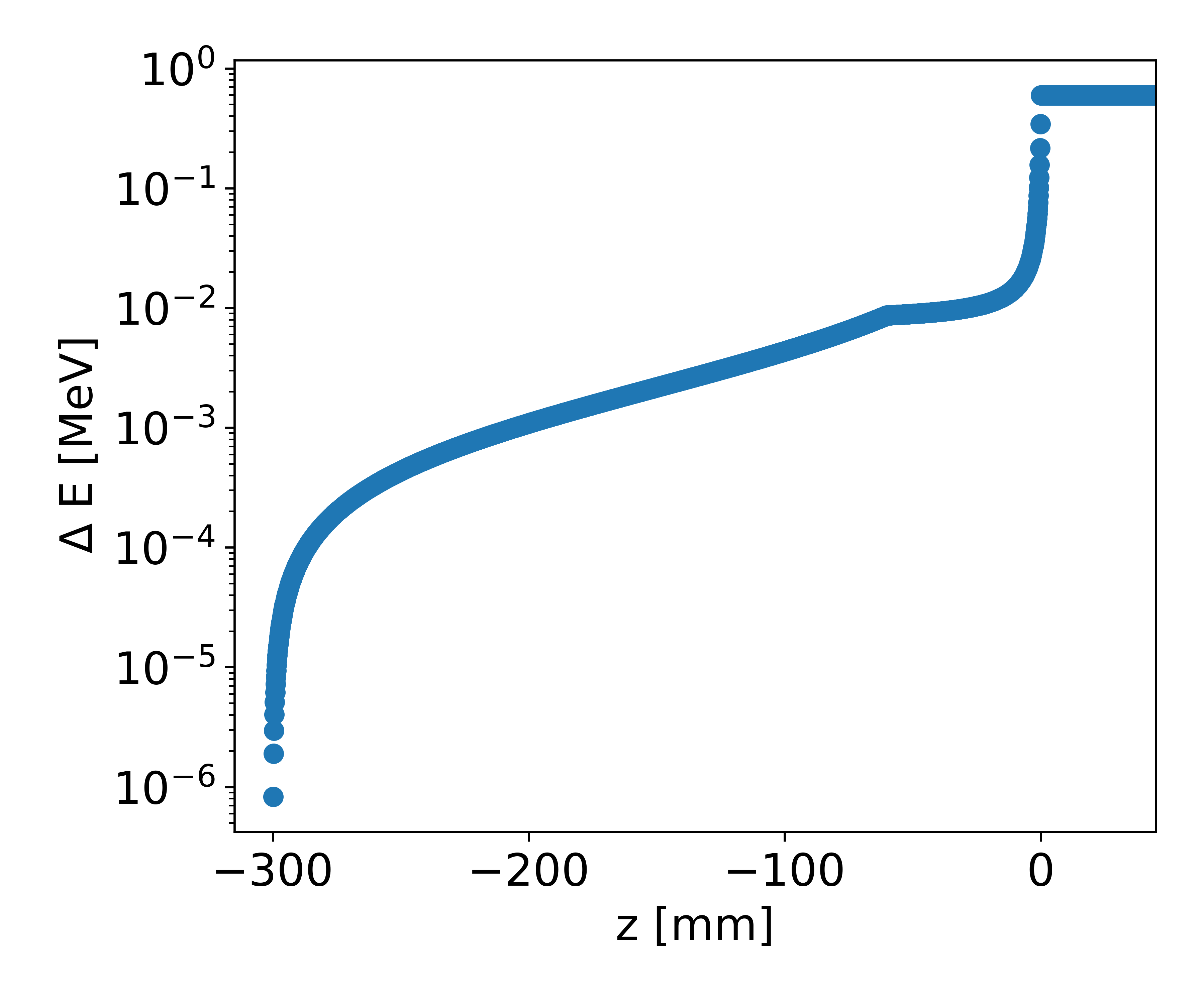}
    \caption{Change in energy of a 35\,MeV electron passing through a pinhole aperture (1\,$\mu$m) in a flat, perfectly conducting surface. Cavity length is 300~mm with the exit aperture positioned at z=0.}
\label{fig:e_dep_35_MeV_elec}
\end{figure}

\subsection{Power radiated toward an exit aperture}
The basic model simulated for this second test is a single charged particle approaching a conducting surface; this surface is treated as a single image charge so that reflected fields may be easily simulated. As mentioned above, we follow this approach with the caveat that we must assume the surface to be a perfectly reflecting conductor or superconducting film.

This model is simple from a computational standpoint because it can be easily adapted for estimating the change in energy for a particle approaching a flat conducting surface and passing through a pinhole aperture or wider apertures in the surface. It also allows for the simulation of a particle's trajectory after the neutralization of the image charge (i.e. the retraction or screening via dielectric of the conducting surface) such that the test particle only interacts afterwards with residual fields. 

A critical illustration of the head-on field dependence is the steep increase in $\Delta E$ for the particle approaching an exit aperture, as shown in Fig.~\ref{fig:e_dep_35_MeV_elec}. Here, the use of a pinhole aperture on the micron scale gives a rough limit for the maximum energy shift due to reflected fields. In this case, the field from the image charge is simply cut off at a radius of $R\simeq a$ where $R$ is the source--image distance and $a$ is the aperture radius.

\begin{figure}[ht!]
    \centering
    \includegraphics[width=0.84\textwidth,trim={0 0.7cm 0 0.5cm},clip]{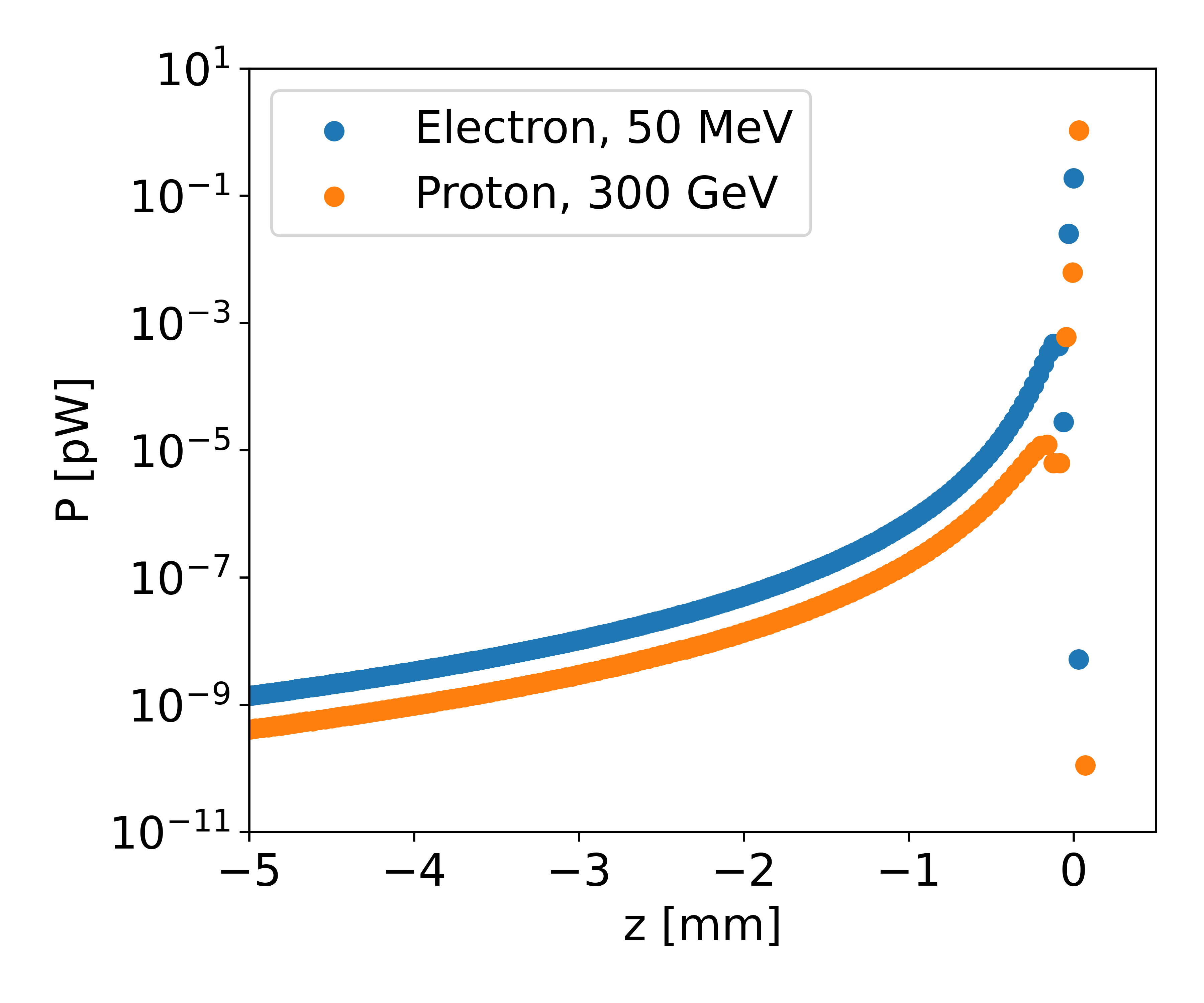}
    \caption{Average radiated power at each integration step for an electron and proton following Eq.~(\ref{eq:lw_power}) (the cross product is ignored). Cavity length is 300~mm, with a 1\,$\mu$m pinhole aperture positioned at $z=0$.}
\label{fig:pow_dep_elec_prot}
\end{figure}

In Fig.~\ref{fig:pow_dep_elec_prot} radiated power is calculated for two similar pinhole-aperture simulations with a proton at 300~GeV and an electron at 50~MeV. The full expression for radiated power from these particles is
\begin{equation}
P=\frac{2}{3} \frac{e^2}{c} \gamma^6\left[(\dot{\boldsymbol{\beta}})^2-(\boldsymbol{\beta} \times \dot{\boldsymbol{\beta}})^2\right]
\label{eq:lw_power}
\end{equation}

\noindent which reduces to the familiar Larmor equation for nonrelativistic velocities. Although the peak power deposition shown here scales to the mW range for bunch populations of $10^9$, the effect diminishes for larger apertures.

\begin{figure}[ht!]
    \centering
    \includegraphics[width=0.84\textwidth,trim={0 0.7cm 0 0.5cm},clip]{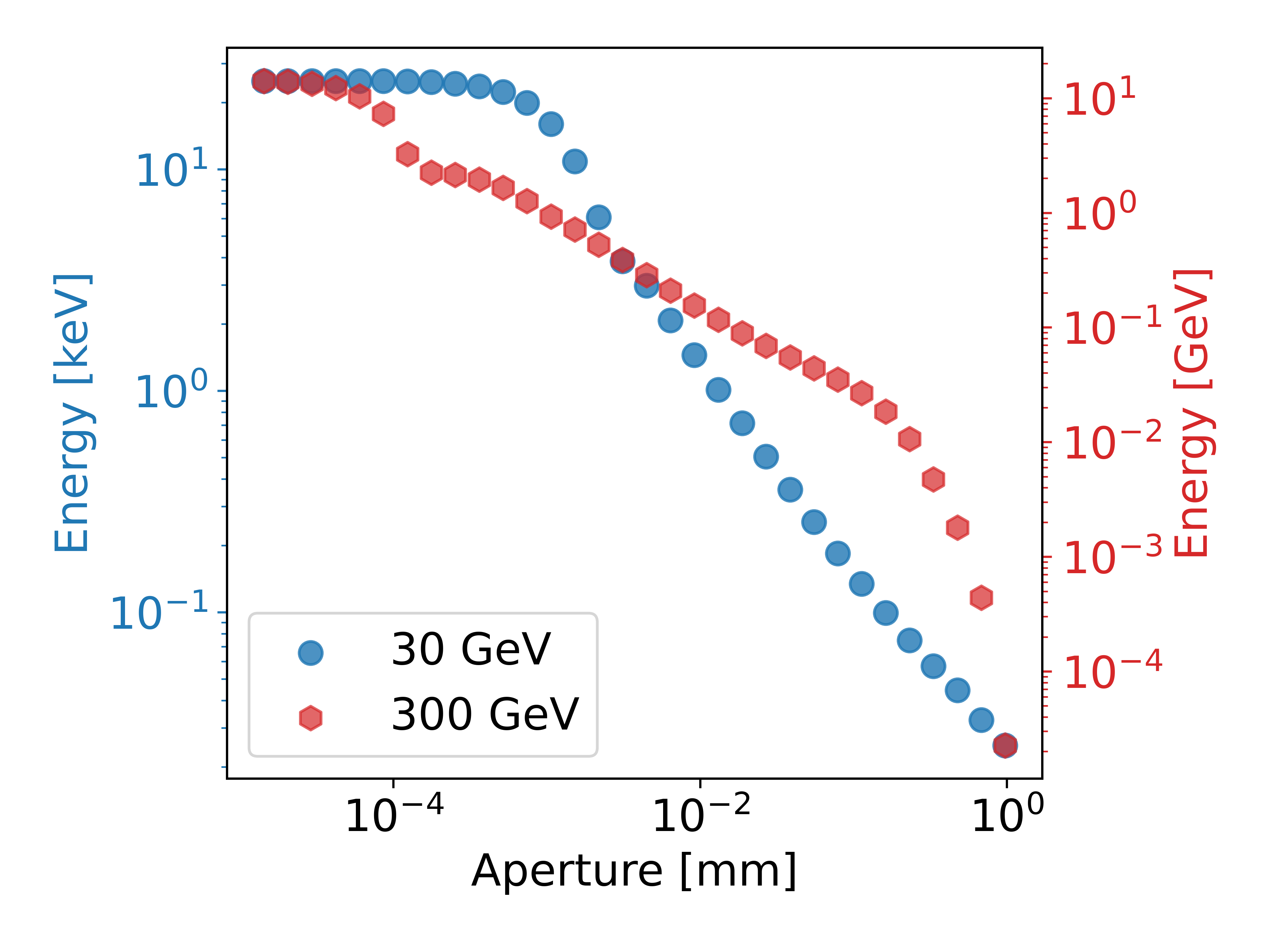}
    \caption{Total energy increase for various exit-aperture sizes for protons traversing a 1\,m cavity with starting energies of 30\,GeV (blue) and 300\,GeV (red). Apertures are set in a perfectly conducting surface.}
\label{fig:e_dep_apertures}
\end{figure}

To study more realistic aperture sizes, the previous image-charge model is no longer accurate once the distance between the incoming particle and the exit aperture reduces to the roughly the size of the aperture radius. To correct this, a transverse shift is added to the image particle's position, corresponding to the solid angle defined by the source--image distance $R$. The shift can be defined as $\Delta x = R \operatorname{tan}\theta$, where $\theta$ is the apex angle related to the solid angle by $\Omega = 2\pi (1-cos\theta)$. Then, to ensure that the empty space of the aperture itself does not contribute any charge when calculating the reflected potentials, we adjust the image charge magnitude to the available solid-angle fraction of reflected charge:

\begin{equation}
    e_{img} = e\left( 1- \frac{2 a^2}{R^2\left[1 - \operatorname{cos}(\pi / 2)\right]} \right)
    \label{eq:image_charge_fraction}
\end{equation}

Results for total energy increase for protons passing through a variety of apertures using these approximations are shown in Fig.~\ref{fig:e_dep_apertures}. The protons simulated here are positioned on the beam axis and have negligible keV-scale starting transverse momentum. Each point on the plot represents a single run through a 1\,m cavity, with an acceleration curve similar to the one shown in Fig~\ref{fig:e_dep_35_MeV_elec}.

From these variable-aperture results, we can see that the energy shift is roughly 5\% for the 300\,GeV protons and essentially negligible for 30\,GeV ones, thus providing additional numerical verification for the asymptotic $\beta$ dependence predicted analytically by Eq.~(\ref{eq:E_n_headon}) and illustrated in Fig.~\ref{fig:rel_lortz_force_both}. In this case, the minimum-aperture values from these simulations converge toward an upper limit on the possible energy increase. It is also worthwhile to note the sharp dropoff in energy exchange for apertures larger than roughly 100\,nm at high energies -- in the next section we will draw on this result as a basis for modeling the rapid withdrawal of conducting surfaces.

\subsection{Approaching a conducting iris or chopper}\label{sec:results_conducting_surface}
Here we illustrate screened-source recoil reduction, which was discussed conceptually in Section~\ref{sec:recoil_reduct_pedagog}. In this case we simulate a more simplified scenario in which the test particle approaches a flat conducting surface that is instantaneously removed from line of sight. This model is intended as a first-order approximation for opening an arbitrarily large aperture in the surface over the course of the simulation.

As mentioned above, we justify the use of this model based on the results shown in Fig.~\ref{fig:e_dep_apertures} which indicate a strong reduction in energy exchange as apertures exceed 100\,nm.\footnote{
For example, if we increase the aperture size by roughly 100\,nm in 1\,ns (implying a chopper speed of 100 m/s) we can assume a corresponding dropoff in reflected energy of roughly an order of magnitude for, e.g., the 300\,GeV protons shown in Fig.~\ref{fig:e_dep_apertures} as they traverse a span along the beam axis of approximately 300\,mm.} Thus, by removing the conducting surface in a single timestep, we are coarsely simulating the dynamics involving the critical distance at which the solid angle of a particle's field distribution no longer collides with the conducting surface around an aperture.

We accomplish this by tracking an image charge of the test particle with the appropriate retardation parameters and then eliminating it at a selected time step.\footnote{An alternative model, which may be more easily implemented in experiment, would use a series of semiconducting plates with fixed-radius apertures instead of mechanical choppers. With this setup, changing from conducting to dielectric states would also allow for low-recoil acceleration of a test particle, albeit with a penalty due to the fixed aperture preventing a significant fraction of the particle's fields from being reflected while in the conducting state and added complexity due to surface-charge effects.} Once the emitted fields from the test charge no longer impinge on the conductor's surface, the charge is left accelerating in a residual field. In other words, the energy shift of the test charge after this cutoff point is solely due to the fields which were reflected from the conductor before its withdrawal.

\begin{figure}[htp!]
    \centering
    \includegraphics[width=0.68\textwidth,trim={0 0.7cm 0 0.5cm},clip]{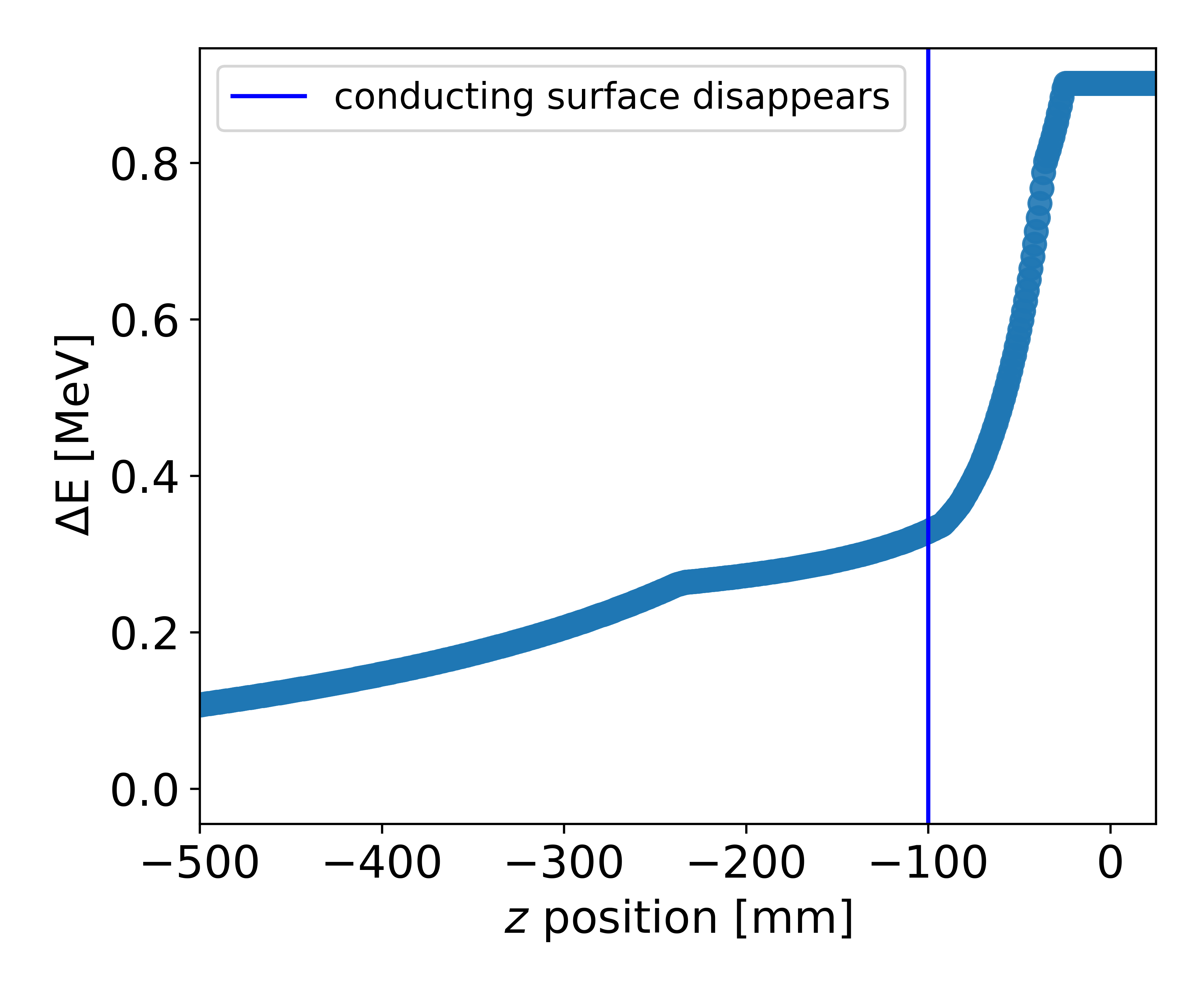}
    \caption{Residual-field kinetic energy gain for a single 85~MeV electron traversing a 1000\,mm cavity toward a conductive iris or chopper. The $y$-axis shows energy increase in MeV from the initial value. The vertical line indicates the integration step where the iris is opened.}
\label{fig:mom_switch_85MeV_elec}
\end{figure}

A test case is shown in Fig.~\ref{fig:mom_switch_85MeV_elec} for an electron with a starting energy of 85\,MeV, where an aperture is opened at 100\,mm from impact. We should emphasize that this is a phenomenon uniquely suited to retarded-potential analysis: for a static integrator, the test particle's acceleration would immediately cease. The momentum change after this cut-off is 0.6\,MeV/c over a span of 220\,ps, indicating an average force of roughly 200\,pN. This constitutes an uncompensated $+z$ momentum shift carried by the test particle to be absorbed downstream by a target, beam stop, or other machine element. When scaled to a bunch population of, e.g., 1 billion particles, this force is on the order of 0.1\,N. 


In general, this test case verifies that an integrator based on Eq.~(\ref{EOM_mom_nospin}) can be used to track residual fields (i.e. tracking a test particle's properly correlated retarded interactions with fields from its respective image charge after a timestep in which the image charge has been withdrawn) with the noteworthy result of an electron being accelerated by its own wake by roughly 1\% of its initial kinetic energy in a 1\,m cavity, where more than half of the acceleration is solely due to residual fields in the final 100\,mm. Although the scale of this acceleration is low in terms of typical accelerating field gradients, it should be emphasized that this is a passive effect involving no external fields. Moreover, since the acceleration of the test particle due to the residual fields causes no corresponding counter-force in the $-\hat{z}$ direction (with the exception of radiation pressure, which, as verified above, is orders of magnitude smaller in this energy range) this form of passive acceleration effectively minimizes drag on the surrounding structure. 

It could be assumed equivalently that the chopper is not opened at all but instead blocked by a rapidly inserted dielectric surface having high electronic stopping power to maximize momentum transfer~\cite{datz_stoppingpow_1996,lindhard_stopping_1996}. In this case, the particle deposits the majority of its momentum into the dielectric surface, and any acceleration which had occurred beyond the opening/switching point will again result in recoil-less deposition of momentum in the $+z$ direction.

\begin{figure}[ht!]
    \centering
    \includegraphics[width=0.84\textwidth,trim={0 0.7cm 0 0.5cm},clip]{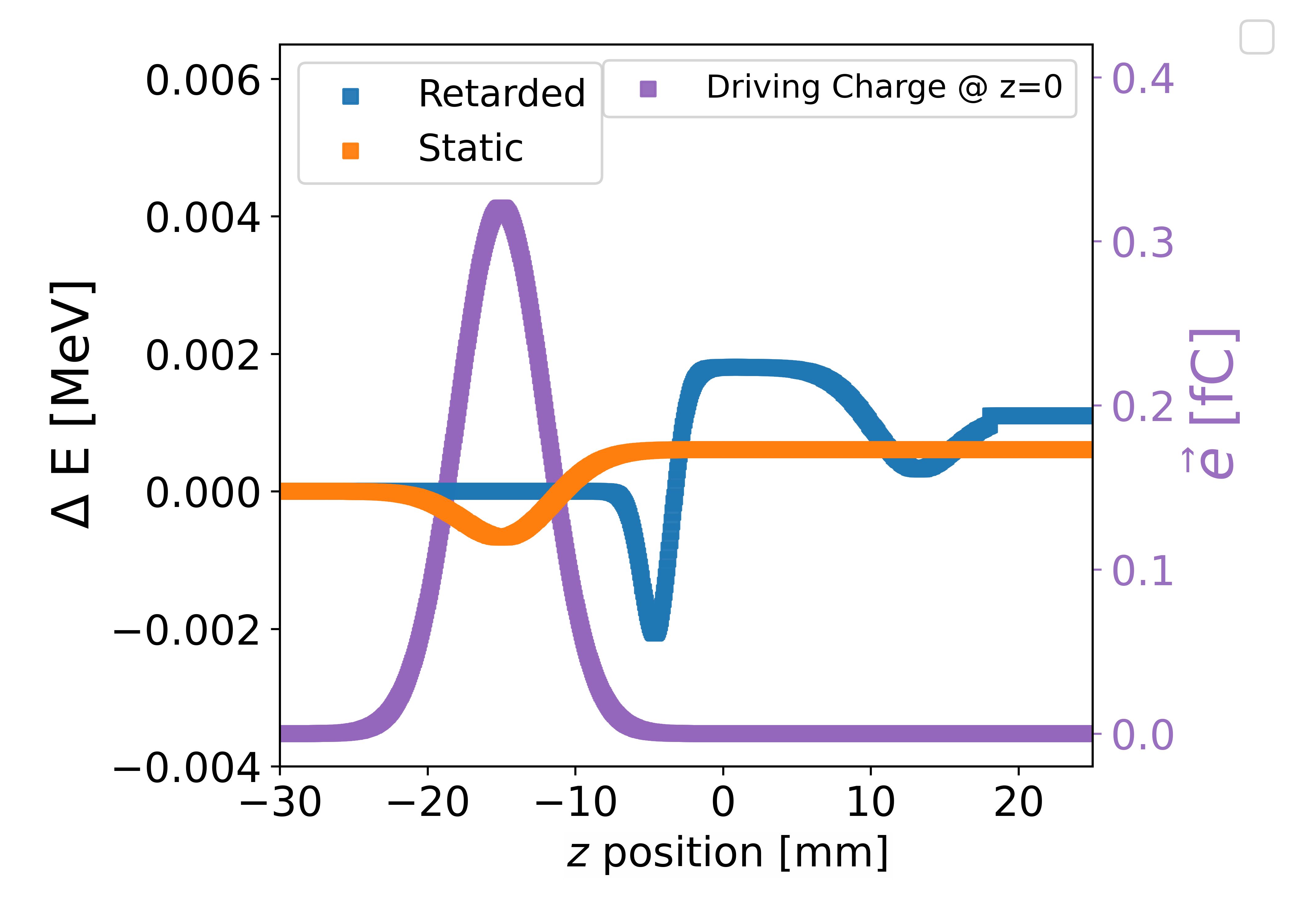}
    \caption{ Energy shift of a 2\,GeV proton traversing the wake of a heavy-ion bunch (scale on left axis). Total charge for the heavy-ion bunch relative to the proton's position (scale on right axis).}
\label{fig:real_bunch_accel_stat_ret}
\end{figure}

\begin{figure}[ht!]
    \centering
    \includegraphics[width=0.84\textwidth,trim={0 0.7cm 0 0.5cm},clip]{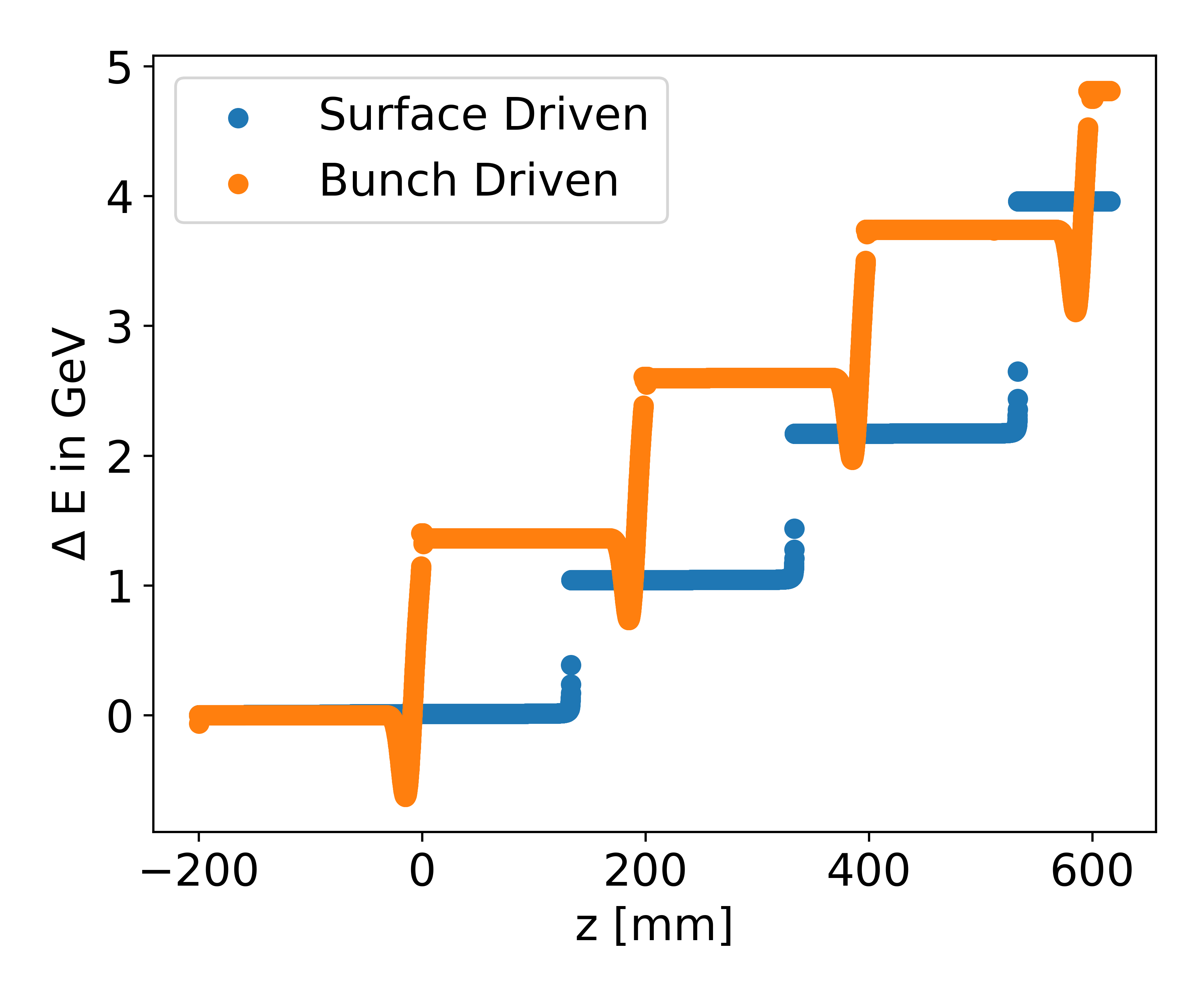}
    \caption{Acceleration of 200\,GeV protons from cascading series of heavy-ion driving bunches (as in Fig.~\ref{fig:real_bunch_accel_stat_ret}) and conductive-surface choppers (as in Fig.~\ref{fig:mom_switch_85MeV_elec}).}
\label{fig:switching_v_realbunch_repeatingcav_200GeV}
\end{figure}

\subsection{Residual wake acceleration from a low-$\beta$ heavy ion bunch}\label{subsec:residual_wake_bunch}

In order produce the screening effect without using mechanical apertures or choppers, we consider a test particle approaching a fixed aperture in a dielectric surface (i.e. a test particle traversing a dielectric collimator). A low-$\beta$, highly-populated driving bunch of heavy ions is then synchronized to cross transversely to the beam axis, on the downstream side of the exit aperture. With this arrangement, transverse forces on the test particle are minimized and it still undergoes residual-field acceleration for a duration proportional to $R/c$ (after the time step where the driving bunch has traversed the full diameter of the aperture and thus passes from line of sight).

Despite its low velocity, the driving bunch is sufficient to provide a short accelerating pulse to the test particle. To simplify the dynamics, a single pair or several pairs of driving bunches crossing from opposite directions on a transverse axis could be then used to negate any transverse force on the test particle.

For the present study, we further simplify the dynamics of the driving bunch, treating it as a time-dependent, Gaussian charge distribution which is synchronized to have a maximum at some arbitrary $R$ value. Then, instead of giving it a transverse velocity, the bunch is placed at rest at the position $\textrm{z}=\textrm{x}=\textrm{y}=0$. In this way, when its charge value falls to zero, the bunch no longer interacts with the incoming test particle. 

Results for a single bunch crossing with this model are shown in Fig.~\ref{fig:real_bunch_accel_stat_ret}. We also show results using a static (i.e. non-retarded) version of Eqs.~\ref{eq:eoms_mom_trans}--\ref{eq:eoms_pos} for comparison. 
Here, the driving bunch is modeled as a macroparticle consisting of gold hexafluoride ions with a mass of 315\,amu and a charge of $-2$, which reaches a peak total charge when the test proton arrives at $\textrm{z}=-20$\,mm (the peak bunch population is $10^9$ or, in terms of total charge, roughly 0.33\,fC). It should be emphasized that the charge is being artificially increased and decreased here to simulate the particles of the driving bunch becoming visible and then blocked due to the obstructing dielectric aperture wall. 

To compare with the model from Section~\ref{sec:results_conducting_surface}, Fig.~\ref{fig:switching_v_realbunch_repeatingcav_200GeV} simulates single 200~GeV protons passing through several cycles of a bunch-driven openings versus a series of conducting-surface choppers. In this case, the driving bunches are identical in mass and charge to those in the previous example, but are placed at 200\,mm intervals starting at $z=0$ and are synchronized to reach a maximum population of $10^8$ when the incoming proton is 20\,mm upstream. For the surface-driven test, the choppers are synchronized to open when the incoming proton is 50\,mm upstream.

The magnitude of the acceleration depends heavily on the test particle's location when the chopper is opened for the conducting-surface model, and when the real-bunch charge density is maximum for bunch-driven model. Thus, in practice, the performance of these methods would depend largely on the attainable precision in terms of chopper speed and driving-bunch synchronization. Nevertheless, the resulting acceleration gradients of roughly 10~GeV/m are considerable, especially given that they are driven by effectively static charges and reflected fields for the bunch and surface tests, respectively (and not requiring, e.g., an external RF field source or driving plasma).

\subsection{Computational and practical considerations}
It is important to note that the conducting surface model is essentially a two-particle system, whereas the real-bunch model uses a well-aligned macroparticle. Preliminary multiparticle tests have also been performed with the conducting surface approach using tens of particles and a variety of transverse spot sizes; these so far have not shown significant deviation from the single-particle simulations, likely due to the highly forward-biased potentials radiating from the test particles. However, multiparticle simulations with a higher particle count using a retarded integrator are highly CPU intensive, since each integration step for a test particle requires pairwise distance calculations for all image charges and/or driving charges in the system (i.e. for the conducting surface model, this includes a single test particle's own image charge and all other test particles' image charges). Because of this, simulations for more than 1000 particles are unfeasible on ordinary workstations. Further multiparticle simulation would thus require cluster computing resources, and may benefit from pairwise-distance calculation optimization (see~\cite[pp.~62--69]{folsom_thesis}). 

Additionally, some screening may occur between leading and trailing particles in a test bunch, with a leading particle's acceleration away from trailing particles in the bunch creating an additional wake (this is an intrabunch effect, \textit{not} involving image charges or a driving bunch). However, one can check using Eq.~(\ref{LW_forces_jackson}) that for receding particles (i.e. $\boldsymbol{\beta}\cdot\boldsymbol{n}\approx-1$) the resulting forces are highly attenuated, being proportional to the Coulomb force times $1/(4\gamma^2)$.

Regardless of computational methods, an overall caveat to these models is that the strong $1/R^2$ dependence (where $R$ is the test particle to source/image particle distance) is likely to lead to significant longitudinal energy spread, with leading particles in a test bunch undergoing much stronger acceleration than trailing one with respect to an oncoming conducting surface or bunch. To address this issue, beam-shaping methods may be necessary, such as using bunches with a transverse spot size that widens from front to rear longitudinally in order to increase the trailing particles' exposure to the conducting surface or driving-bunch wake; using a hollowed or heavily haloed driving-bunch distribution; or by microbunching of the test-particle beam.

For the bunch-driven case, we must also assume that the results shown here are over-optimistic in terms of the driving bunch's transverse size, since only a single macroparticle of $10^8$--$10^9$ particles was used with a random 10\,{$\mu$}m scale transverse alignment error, and interaction with the high-$\beta$ test particle falls off strongly for test particle--driver bunch misalignments beyond this range.

Thus, in practice, a large driving-bunch population would need to occupy a small, micron-scale transverse footprint with a population and transverse velocity corresponding to the charge-density curve in Fig.~\ref{fig:real_bunch_accel_stat_ret} in order to achieve the energy shifts shown therein. The method mentioned above of using several pairs of driving bunches crossing on antiparallel trajectories transverse to the beam axis may help to accomplish this (i.e. minimizing space charge until the instant all pairs of driving bunches reach the beam axis).
We should also note that the conducting-surface method does not suffer from this drawback, since the test particle's image charges remain transversely aligned until the conductor is withdrawn or screened.

In general, there is a tradeoff between the two models simulated here: in the surface-driven case, the test particle witnesses a high-$\beta$ image charge, leading to a $\vec{e}\,\cev{e}\gamma^5 / R^2$ proportionality in the $1/(\kappa^2 R^2)$ dependent terms for calculating conjugate momentum shown in Eqs.~(\ref{eq:eoms_mom_trans}) and (\ref{eq:eoms_mom_long});\,\footnote{This can derived from Eq.~(\ref{EOM_mom_nospin}) for the image-charge scenario by first rationalizing so that $1/\kappa = \vec{\gamma}^2 (1+\vec{\beta}) \approx 2\vec{\gamma}^2$, then taking $\cev{\beta}\cdot\vec{\beta}=|\beta|^2 cos(\theta) \approx -|\beta|^2$, and $\cev{\gamma}\approx\vec{\gamma}$.} whereas in the bunch-driven case, the driving particles have a negligible longitudinal velocity, thus $\vec{\gamma}\approx 1$ and must be compensated by a high bunch population. In other words, in a driving-bunch particle's integration steps, it is accelerated by an $\cev{e}\vec{e}\,\cev{\gamma}^4 / R^2$ dependent conjugate momentum from the test particle; but the test-particle is only accelerated in response by a $\cev{e}\vec{e}\,\cev{\gamma} / R^2$ conjugate momentum.


\subsection{Test case: comparison with dielectric laser micro-acceleration}
As a final example, we draw from recent advances in micro-scale dielectric laser accelerators (DLAs) where an inverse Cherenkov radiation (ICR) effect has shown promising results~\cite{liu_2020,sun_bin_2023,sun_li_2023,volkov_2022,Sun_li_2021,Kozak_17}. In its simplest implementation, this scheme uses a dielectric prism with its hypotenuse edge aligned parallel to the axis of an accelerating electron beam. A laser pulse is directed through an upper surface of the prism such that induced Cherenkov radiation in the prism is totally reflected away from the beam axis. The phase velocity of the resulting surface waves which \textit{do} reach the beam axis can be synchronized with an incoming electron beam. This arrangement results in a cancellation of the Lorentz force of these surface waves' transverse magnetic and electric components, leaving a single longitudinal accelerating field, $E_z$.

This ICR-DLA type structure is well-suited for comparison with the methods of the present work, since the dielectric surroundings eliminate the need to consider transverse beam effects such as beam loading. In particular, we draw from \cite{liu_2020}, which simulated 10\,MeV electrons accelerating through a quartz ICR-DLA for an energy gain of $\Delta E = 0.3$\,MeV over a span of 200\,{${\mu}$}m; the electric field intensity of the incident laser was 5\,GV/m and the magnitude of the resulting longitudinal surface field $E_z$ oscillated around a value of approximately $-1.5$\,GV/m.

\begin{figure}
  \begin{subfigure}{\columnwidth}
    \centering
    \includegraphics[width=0.75\linewidth,trim={0 0.7cm 0 0.5cm},clip]{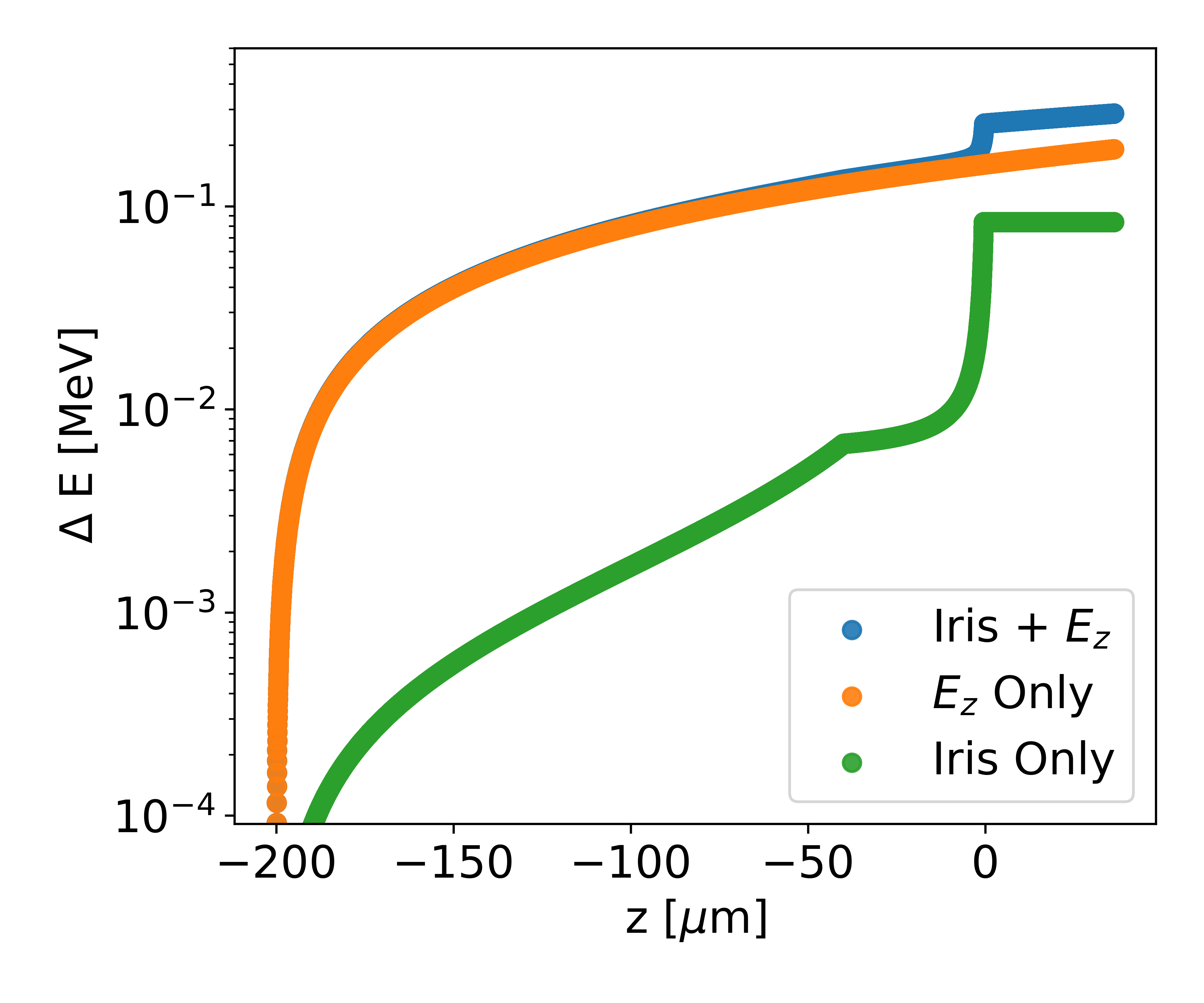}
    \caption{}
    \label{fig:external_DLA_vs_iris_single}
  \end{subfigure}
  
  \begin{subfigure}{\columnwidth}
    \centering
    \includegraphics[width=0.75\linewidth,trim={0 0.7cm 0 0.5cm},clip]{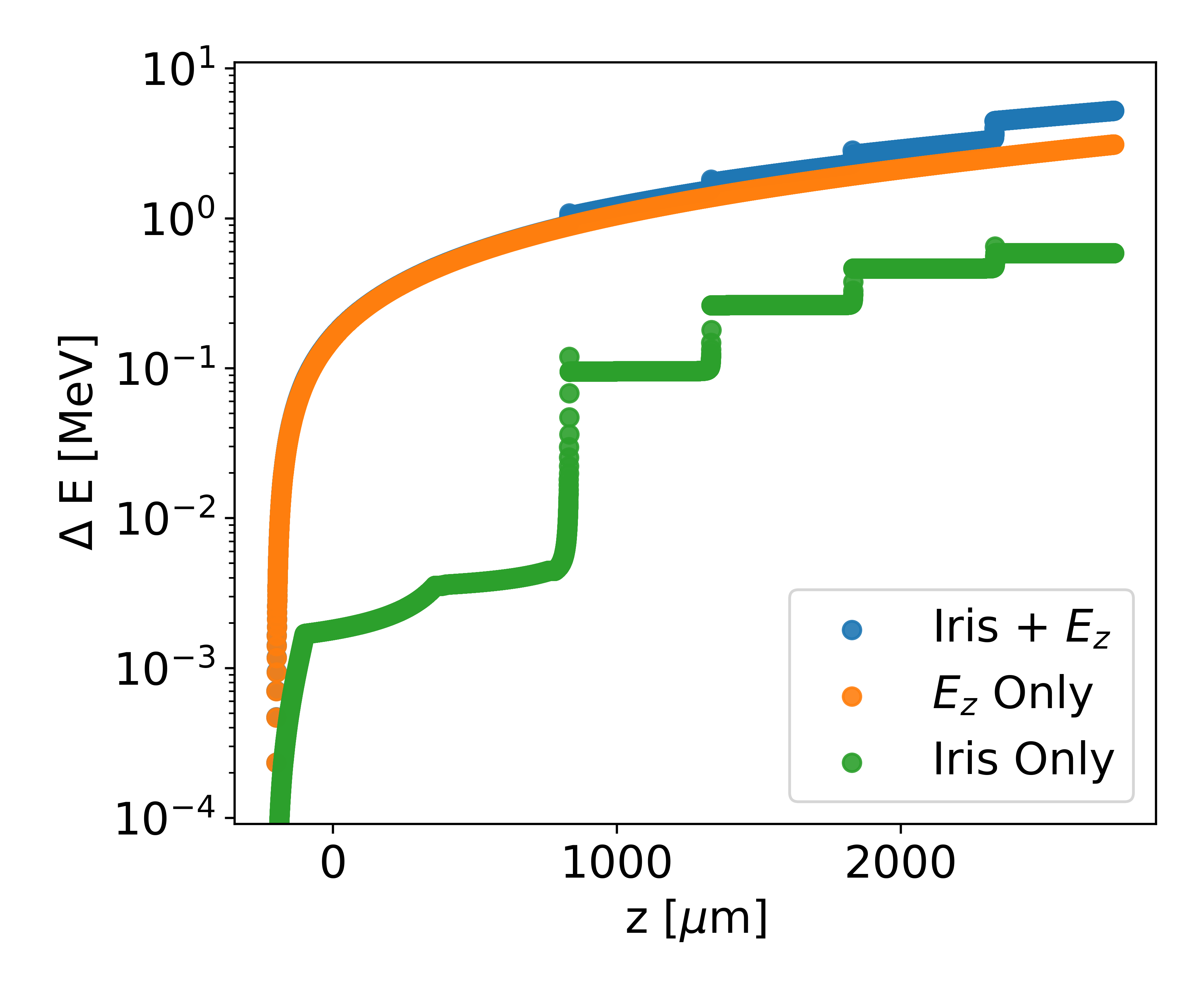}
    \caption{}
    \label{fig:external_DLA_vs_iris_repeat}
  \end{subfigure}
  \vspace{-1\baselineskip}
  \caption{Energy shift comparison and combination with an external field $E_z$ of $-1.5$\,GV/m for 10\,MeV electrons traversing an (a)~200\,{${\mu}$}m cavity having a single iris opening at $\Delta z = 1\,\mu$m and (b) a 2.75\,mm cavity having a series of irises opening at $\Delta z = 100\,\mu$m.}
  \label{fig:external_DLA_vs_iris}
\end{figure}

For the results shown in Fig.~\ref{fig:external_DLA_vs_iris_single}, we simulated a simplified version of such an ICR-DLA, taking its surface field $E_z$ as a constant $-1.5$\,GV/m, where an energy gain of roughly 0.1\,MeV is observed. In covariant terms, this can be added as a time-independent field to Eq.~(\ref{eq:eoms_mom_long}) as follows
\begin{align}\label{eq:conjmom_ext_field}
    \frac{d \mathcal{P}^z_{ext}}{d \tau} &= \frac{e}{c}V_{\beta}\partial^z A^0_{ext}
    \nonumber \\
    &=  e\gamma\left(1-\boldsymbol{\beta}\right)\partial^z A^0_{ext}
    \nonumber \\
    &=  e\gamma\left(1-\boldsymbol{\beta}\right)E_z
\end{align}
\noindent where the \textit{ext} subscript indicates an externally applied potential. Here, we have used the definition $E_z = \partial^z A^0_{ext} - \partial^0 A^z_{ext}$ where the time derivative is zero. The results for a test electron traversing this external potential (which are in good agreement with the results from \cite{liu_2020} are compared with one approaching a retracting iris, similar to the result shown in Fig.~\ref{fig:mom_switch_85MeV_elec}. We also show results for a case which combines both the retracting iris and the ICR-DLA in a single cavity.

For this simulation, the iris is located at $z=0$ and is opened instantaneously when the incoming electron reaches $z=-1$\,${\mu}$m. At this scale, the conducting-surface acceleration is negligible once the iris has opened by roughly 50\,nm, yet the actuation speed of such an iris or chopper may be unrealistically high in practice (i.e. retracting a conducting surface a distance of 50\,nm in the span of roughly 3.3\,fs, or at a speed of roughly 0.05$c$).

In Fig.~\ref{fig:external_DLA_vs_iris_repeat}, we consider instead a cascading series of irises spaced 500\,{$\mu$}m apart along a 2.75\,mm ICR-DLA device (or series of devices) and with an initial electron position of $z=-200\,\mu$m. Here, each iris opens when the incoming electron beam is at a distance of 100\,{$\mu$}m, resulting in an approximate factor of two gain versus the $E_z$-only trial.

Although we have elected to show results using the conductive surface model for its computational simplicity, the driving bunch model may in fact be more feasible on such a scale, since it only requires synchronizing a series of low-$\beta$ bunches passing transversely along the downstream side of a dielectric aperture, instead of requiring high-frequency conductive irises or choppers.

While these simulation results are very preliminary, it is worth observing here that an the effects of external field can be incorporated into a covariant integrator's equations of motion using the relatively simple Eq.~(\ref{eq:conjmom_ext_field}) for good agreement with the more conventional simulation methods.

\section{Conclusion}
We have demonstrated a retarded, covariant simulation framework for testing power deposition and acceleration of high-$\beta$ protons and electrons traveling toward conducting surfaces and passing though a variety of apertures which were either fixed-radius or allowed to open instantaneously. We also simulated acceleration due to residual fields from either conducting surfaces or external charges after breaking line of sight with the test particle.

Analytical predictions for a $\gamma^2$ proportionality of the Lorentz-force in highly aligned near-collisions between charged particles were validated numerically. Specifically, the stopping power of individual protons in near-collisions with heavy ions was shown to increase dramatically in the 1--4\,TeV range.

The deposited power from single electrons and protons passing a micron-scale pinhole aperture is shown to reach the 10\,pW scale, driven solely by the particles' reflected fields. However, as apertures are opened beyond a radius of roughly $1\,\mu$m, there is a sharp decrease in deposited power, thanks to the narrowly forward-biased shape of the incoming particles' fields. 

An acceleration of roughly 1\% is observed for 85~MeV electrons approaching a disappearing conducting surface (i.e. an iris or chopper), with the majority of acceleration occurring after the surface has withdrawn. Although this result is noteworthy, in practice it will likely be more complicated when accounting for surface roughness, surface charge accumulation, chopper speeds, etc. The results for a transversely crossing external bunch are shown to have comparable performance with the surface-driven acceleration (for driving-bunch populations upwards of $10^8$). In this case, the distance between the test particle and driving bunch is critical.

A resulting longitudinal energy spread is a fundamental issue for either model, thanks to the leading test particles witnessing stronger $1/R^2$ dependent fields than trailing ones. Beam shaping or microbunching may help mitigate this issue. Another general concern with the simulation methods presented here is that it is difficult to run multiparticle retarded-potential simulations with populations beyond hundreds of particles; this is due to an $N^2$ scaling of pairwise distance calculations.

A comparison was also made between the conducting surface model and the ICR-DLA micro-acceleration technique, where we treated the accelerating gradient from an ICR-DLA device as a static $E$ field. When combining such a device with a series of conducting irises over a length of 2\,mm, we observed a factor of two gain in energy shift versus a solitary ICR-DLA.


For the screened-source cases where the conducting iris or chopper is opened or the driving bunch is blocked from line of sight at large $R$, the test particle continues to accelerate from residual fields without having a charged or conducting body to interact with; it thus causes no countervailing force in the $-z$ direction. In other words, no recoil occurs on the encompassing structure beyond the opening/screening instant. This phenomenon may prove particularly useful for reducing structural vibrations in, for example, micro-accelerator applications.

The concepts of reflected self-wake acceleration and screened-source wake acceleration are dependent upon retarded-potential analysis, and may merit further study in the context of space-limited or high-precision experiments. More generally, the improvement in accuracy over standard beam physics software demonstrated here for high energy single-particle interactions may also be useful for improved modeling of, for example, pre-collision dynamics, electron lensing, or other beam--beam effects. 


 \bibliographystyle{elsarticle-num} 
 \bibliography{cas-refs}

\begin{thebibliography}{10}
\expandafter\ifx\csname url\endcsname\relax
  \def\url#1{\texttt{#1}}\fi
\expandafter\ifx\csname urlprefix\endcsname\relax\def\urlprefix{URL }\fi
\expandafter\ifx\csname href\endcsname\relax
  \def\href#1#2{#2} \def\path#1{#1}\fi

\bibitem{jackson2012classical}
J.~D. Jackson, Classical Electrodynamics, John Wiley \& Sons, 2012.

\bibitem{wolski2014beam}
A.~Wolski, Beam dynamics in high energy particle accelerators, World Scientific, 2014.
\newblock \href {https://doi.org/10.1142/p899} {\path{doi:10.1142/p899}}.

\bibitem{liu_2020}
W.~Liu, Z.~Yu, L.~Sun, Y.~Liu, Q.~Jia, H.~Xu, B.~Sun, Microscale laser-driven particle accelerator using the inverse cherenkov effect, Phys. Rev. Appl. 14 (2020) 014018.
\newblock \href {https://doi.org/10.1103/PhysRevApplied.14.014018} {\path{doi:10.1103/PhysRevApplied.14.014018}}.

\bibitem{folsom:ipac2021-tupab218}
B.~Folsom, E.~Laface, {Fully Covariant Two-Particle Space-Charge Dynamics Using the Liénard-Wiechert Potentials}, in: Proc. IPAC'21, no.~12 in International Particle Accelerator Conference, JACoW Publishing, Geneva, Switzerland, 2021, pp. 1931--1934.
\newblock \href {https://doi.org/10.18429/JACoW-IPAC2021-TUPAB218} {\path{doi:10.18429/JACoW-IPAC2021-TUPAB218}}.

\bibitem{laface_2020_covariant}
E.~Laface, B.~T. Folsom, Covariant hamiltonian approach for time-dependent potentials applied to a pill-box cavity, Phys. Rev. Accel. Beams 23 (2020) 104001.
\newblock \href {https://doi.org/10.1103/PhysRevAccelBeams.23.104001} {\path{doi:10.1103/PhysRevAccelBeams.23.104001}}.

\bibitem{bordovitsyn_technique_2003}
V.~A. Bordovitsyn, T.~O. Pozdeeva, The {Technique} of {Covariant} {Differentiation} of {Retarded} {Potentials}, Russian Physics Journal 46~(5) (2003) 457--469.
\newblock \href {https://doi.org/10.1023/A:1026217822337} {\path{doi:10.1023/A:1026217822337}}.

\bibitem{barut1980electrodynamics}
A.~O. Barut, \href{http://pi.lib.uchicago.edu/1001/cat/bib/502533}{Electrodynamics and classical theory of fields \& particles}, Courier Corporation, 1980.
\newline\urlprefix\url{http://pi.lib.uchicago.edu/1001/cat/bib/502533}

\bibitem{Ryne:IPAC2018-THPAK044}
R.~Ryne, B.~Carlsten, C.~Mitchell, J.~Qiang, {S}elf{-C}onsistent {M}odeling using a {L}ienard{-W}iechert {P}article{-M}esh {M}ethod, in: Proc. 9th International Particle Accelerator Conference (IPAC'18), Vancouver, BC, Canada, April 29-May 4, 2018, no.~9 in International Particle Accelerator Conference, JACoW Publishing, Geneva, Switzerland, 2018, pp. 3313--3315.
\newblock \href {https://doi.org/doi:10.18429/JACoW-IPAC2018-THPAK044} {\path{doi:doi:10.18429/JACoW-IPAC2018-THPAK044}}.

\bibitem{ryne:ipac12-tuppp036}
R.~D. Ryne, C.~E. Mitchell, J.~Qiang, B.~E. Carlsten, N.~A. Yampolsky, \href{https://jacow.org/IPAC2012/papers/TUPPP036.pdf}{{Large-scale Simulation of Synchrotron Radiation using a Lienard-Wiechert Approach}}, in: Proc. IPAC'12, JACoW Publishing, Geneva, Switzerland, 2012, pp. 1689--1691.
\newline\urlprefix\url{https://jacow.org/IPAC2012/papers/TUPPP036.pdf}

\bibitem{quattromini_LW_2009}
M.~Quattromini, L.~Giannessi, Covariant self-fields regularization in dense electron beams, Il Nuovo Cimento C 32~(2) (2009) 169--172.
\newblock \href {https://doi.org/10.1393/ncc/i2009-10407-7} {\path{doi:10.1393/ncc/i2009-10407-7}}.

\bibitem{medinaRadiationReactionClassical2006}
R.~Medina, Radiation reaction of a classical quasi-rigid extended particle, Journal of Physics A: Mathematical and General 39~(14) (2006) 3801--3816.
\newblock \href {https://doi.org/10.1088/0305-4470/39/14/021} {\path{doi:10.1088/0305-4470/39/14/021}}.

\bibitem{rafelski_2022_rad_reaction}
W.~Price, M.~Formanek, J.~Rafelski, Radiation reaction and limiting acceleration, Phys. Rev. D 105 (2022) 016024.
\newblock \href {https://doi.org/10.1103/PhysRevD.105.016024} {\path{doi:10.1103/PhysRevD.105.016024}}.

\bibitem{rohrlichDynamicsChargedParticle2008}
F.~Rohrlich, Dynamics of a charged particle, Physical Review E 77~(4) (2008) 046609.
\newblock \href {https://doi.org/10.1103/PhysRevE.77.046609} {\path{doi:10.1103/PhysRevE.77.046609}}.

\bibitem{martin-lunaNovelReactionForce2022}
P.~Martín-Luna, D.~González-Iglesias, B.~Gimeno, D.~Esperante, C.~Blanch, N.~Fuster-Martínez, P.~Martinez-Reviriego, J.~Fuster, Novel reaction force for ultra-relativistic dynamics of a classical point charge, arXiv:2208.11096 [physics] (Aug. 2022).
\newblock \href {https://doi.org/10.48550/arXiv.2208.11096} {\path{doi:10.48550/arXiv.2208.11096}}.

\bibitem{griffithsAbrahamLorentzLandau2010a}
D.~J. Griffiths, T.~C. Proctor, D.~F. Schroeter, Abraham–{Lorentz} versus {Landau}–{Lifshitz}, American Journal of Physics 78~(4) (2010) 391--402.
\newblock \href {https://doi.org/10.1119/1.3269900} {\path{doi:10.1119/1.3269900}}.

\bibitem{steaneReducedorderAbrahamLorentzDiracEquation2015}
A.~M. Steane, Reduced-order {Abraham}-{Lorentz}-{Dirac} equation and the consistency of classical electromagnetism, American Journal of Physics 83~(3) (2015) 256--262.
\newblock \href {https://doi.org/10.1119/1.4897951} {\path{doi:10.1119/1.4897951}}.

\bibitem{hammondRadiationReactionUltrahigh2010}
R.~T. Hammond, Radiation reaction at ultrahigh intensities, Physical Review A 81~(6) (2010) 062104, publisher: American Physical Society.
\newblock \href {https://doi.org/10.1103/PhysRevA.81.062104} {\path{doi:10.1103/PhysRevA.81.062104}}.

\bibitem{liInducedCurrentDue2022}
D.~Li, P.~Y. Wong, D.~Chernin, Y.~Y. Lau, Induced {Current} {Due} to {Electromagnetic} {Shock} {Produced} by {Charge} {Impact} on a {Conducting} {Surface}, IEEE Transactions on Plasma Science 50~(9) (2022) 2838--2844.
\newblock \href {https://doi.org/10.1109/TPS.2022.3187667} {\path{doi:10.1109/TPS.2022.3187667}}.

\bibitem{sampathExtremelyDenseGammaRay2021}
A.~Sampath, X.~Davoine, S.~Corde, L.~Gremillet, M.~Gilljohann, M.~Sangal, C.~H. Keitel, R.~Ariniello, J.~Cary, H.~Ekerfelt, et~al., Extremely dense gamma-ray pulses in electron beam-multifoil collisions, Physical Review Letters 126~(6) (2021) 064801.
\newblock \href {https://doi.org/10.1103/PhysRevLett.126.064801} {\path{doi:10.1103/PhysRevLett.126.064801}}.

\bibitem{steinmetzMagneticDipoleMoment2019a}
A.~Steinmetz, M.~Formanek, J.~Rafelski, Magnetic dipole moment in relativistic quantum mechanics, The European Physical Journal A 55~(3), pp. 1--17 (Mar. 2019).
\newblock \href {https://doi.org/10.1140/epja/i2019-12715-5} {\path{doi:10.1140/epja/i2019-12715-5}}.

\bibitem{VAY2018476}
J.-L. Vay, A.~Almgren, J.~Bell, L.~Ge, D.~Grote, M.~Hogan, O.~Kononenko, R.~Lehe, A.~Myers, C.~Ng, et~al., Warp-{X}: A new exascale computing platform for beam–plasma simulations, Nuclear Instruments and Methods in Physics Research Section A: Accelerators, Spectrometers, Detectors and Associated Equipment 909 (2018) 476--479, 3rd European Advanced Accelerator Concepts workshop (EAAC2017).
\newblock \href {https://doi.org/https://doi.org/10.1016/j.nima.2018.01.035} {\path{doi:https://doi.org/10.1016/j.nima.2018.01.035}}.

\bibitem{datz_stoppingpow_1996}
S.~Datz, H.~F. Krause, C.~R. Vane, H.~Knudsen, P.~Grafstr\"om, R.~H. Schuch, Effect of nuclear size on the stopping power of ultrarelativistic heavy ions, Phys. Rev. Lett. 77 (1996) 2925--2928.
\newblock \href {https://doi.org/10.1103/PhysRevLett.77.2925} {\path{doi:10.1103/PhysRevLett.77.2925}}.

\bibitem{lindhard_stopping_1996}
J.~Lindhard, A.~H. S\o{}rensen, Relativistic theory of stopping for heavy ions, Phys. Rev. A 53 (1996) 2443--2456.
\newblock \href {https://doi.org/10.1103/PhysRevA.53.2443} {\path{doi:10.1103/PhysRevA.53.2443}}.

\bibitem{folsom_thesis}
B.~Folsom, \href{https://portal.research.lu.se/en/publications/nonlinear-beam-physics}{Nonlinear Beam Physics}, {L}und {U}niversity, {D}epartment of {P}hysics, 2019, {P}h{D} Thesis.
\newline\urlprefix\url{https://portal.research.lu.se/en/publications/nonlinear-beam-physics}

\bibitem{sun_bin_2023}
B.~Sun, Y.-F. He, R.-Y. Luo, T.-Y. Zhang, Q.~Zhou, S.-Y. Wang, J.~Zheng, Z.-Q. Zhao, On-chip stackable dielectric laser accelerator, Nuclear Science and Techniques 34~(2) (2023) 23.
\newblock \href {https://doi.org/10.1007/s41365-023-01174-7} {\path{doi:10.1007/s41365-023-01174-7}}.

\bibitem{sun_li_2023}
L.~Sun, W.~Liu, L.~Zhang, Q.~Jia, H.~Xu, S.~Liu, High-bunch-charge and high-energy-gain inverse-cherenkov particle accelerator with prebunched beam, IEEE Transactions on Electron Devices 70~(5) (2023) 2514--2520.
\newblock \href {https://doi.org/10.1109/TED.2023.3253099} {\path{doi:10.1109/TED.2023.3253099}}.

\bibitem{volkov_2022}
M.~Volkov, E.~Ramachandran, M.~Mattes, A.~B. Swain, M.~Tsarev, P.~Baum, Spatiotemporal attosecond control of electron pulses via subluminal terahertz waveforms, ACS Photonics 9~(10) (2022) 3225--3233.
\newblock \href {https://doi.org/10.1021/acsphotonics.2c01432} {\path{doi:10.1021/acsphotonics.2c01432}}.

\bibitem{Sun_li_2021}
L.~Sun, W.~Liu, J.~Zhou, Y.~Zhu, Z.~Yu, Y.~Liu, Q.~Jia, B.~Sun, H.~Xu, {GV} m$^{-1}$ on-chip particle accelerator driven by few-cycle femtosecond laser pulse, New Journal of Physics 23~(6) (2021) 063031.
\newblock \href {https://doi.org/10.1088/1367-2630/ac03cf} {\path{doi:10.1088/1367-2630/ac03cf}}.

\bibitem{Kozak_17}
M.~Koz\'{a}k, P.~Beck, H.~Deng, J.~McNeur, N.~Sch\"{o}nenberger, C.~Gaida, F.~Stutzki, M.~Gebhardt, J.~Limpert, A.~Ruehl, et~al., Acceleration of sub-relativistic electrons with an evanescent optical wave at a planar interface, Opt. Express 25~(16) (2017) 19195--19204.
\newblock \href {https://doi.org/10.1364/OE.25.019195} {\path{doi:10.1364/OE.25.019195}}.

\end{thebibliography}





\end{document}